\documentclass[preprint,12pt]{elsarticle}

\usepackage{amssymb}

\usepackage{url}
\usepackage{breakurl}
\usepackage[breaklinks]{hyperref}
\usepackage{xspace}
\usepackage[freestanding]{hepunits}
\usepackage{booktabs}
\usepackage{subcaption}

\DeclareRobustCommand{\chips}{\textsc{Chips}\xspace}

\DeclareRobustCommand{\chipsfive}{\textsc{Chips-5}\xspace}
\DeclareRobustCommand{\numi}{NuMI\xspace}
\DeclareRobustCommand{\nova}{NOvA\xspace}

\usepackage{lipsum}
\makeatletter
\def\ps@pprintTitle{%
 \let\@oddhead\@empty
 \let\@evenhead\@empty
 \def\@oddfoot{}%
 \let\@evenfoot\@oddfoot}
\makeatother

\journal{Nuclear Instruments and Methods in Physics}
\begin{document}
\begin{frontmatter}

\title{Neutrino Characterisation using Convolutional Neural Networks in CHIPS water Cherenkov detectors.}

\author[inst1]{Josh Tingey}
\author[inst1]{Simeon Bash}
\author[inst1]{John Cesar}
\author[inst1]{Thomas Dodwell}
\author[inst1]{Stefano Germani}
\author[inst2]{Paul Kooijman}
\author[inst1]{Petr Mánek}
\author[inst1]{Mustafa Ozkaynak}
\author[inst1]{Andy Perch}
\author[inst1]{Jennifer Thomas}
\author[inst1]{Leigh Whitehead}
\affiliation[inst1]{
    organization={Department of Physics and Astronomy},
    addressline={UCL, Gower St}, 
    city={London},
    postcode={WC1E 6BT},
    country={UK}
}
\affiliation[inst2]{
    organization={Nikhef},
    addressline={Science Park 105, 1098 XG}, 
    city={Amsterdam},
    country={The Netherlands}
}

\begin{abstract}
This work presents a novel approach to water Cherenkov neutrino detector event reconstruction and classification. Three forms of a Convolutional Neural Network have been trained to reject cosmic muon events, classify beam events, and estimate neutrino energies, using only a slightly modified version of the raw detector event as input. When evaluated on a realistic selection of simulated CHIPS-5kton prototype detector events, this new approach significantly increases performance over the standard likelihood-based reconstruction and simple neural network classification.
\end{abstract}



\end{frontmatter}


\section{Introduction}
\label{sec:introduction}
In pursuit of answers to some of the open questions in physics, the study of the phenomenon of neutrino oscillations has become paramount, owing to the implication of the neutrinos' small masses in relation to the mass of the Universe and, potentially, the question of the matter anti-matter asymmetry. The next generation of giant neutrino detectors aim to push the size and therefore detector mass to the 100-500 kilo-ton level. Such massive detectors will quickly become limited by systematic uncertainties, whereas most present experiments are still statistics limited.

For detectors to remain practical and affordable into the future, a novel design strategy is highly desirable in order to break the systematic limitation of single large detectors. One approach is to break the financial dependency, which is presently on the order of \$20M/kilo-ton, such that several identical detectors can be afforded. When such detectors are then placed at different positions in the neutrino beam, the oscillation parameters could be measured with substantial cancellation of the aforementioned systematic errors.

\subsection{The CHIPS R\&D Project}
\label{sec:intro_chips}
The \chips R\&D project aims to develop such a novel strategy: a `cheap as chips' water Cherenkov detector has been designed~\cite{adamson2013} and was built in 2018-2019. Primarily aimed for deployment in long-baseline accelerator beam scenarios, the idea was to lower the cost per kt of sensitive mass to between \$200k-\$300k. For comparison, the Super-Kamiokande detector cost approximately \$4 million per kt to build more than 30 years ago. As physics sensitivity depends on the detector performance in addition to sensitive mass, this comparison is not entirely rigorous; however, it highlights the scale of possible cost savings. This paper focusses on the simulation and reconstruction of events in the 5~kton \chips detector (\chipsfive).

\begin{figure} 
    \centering
    \includegraphics[width=0.6\textwidth]{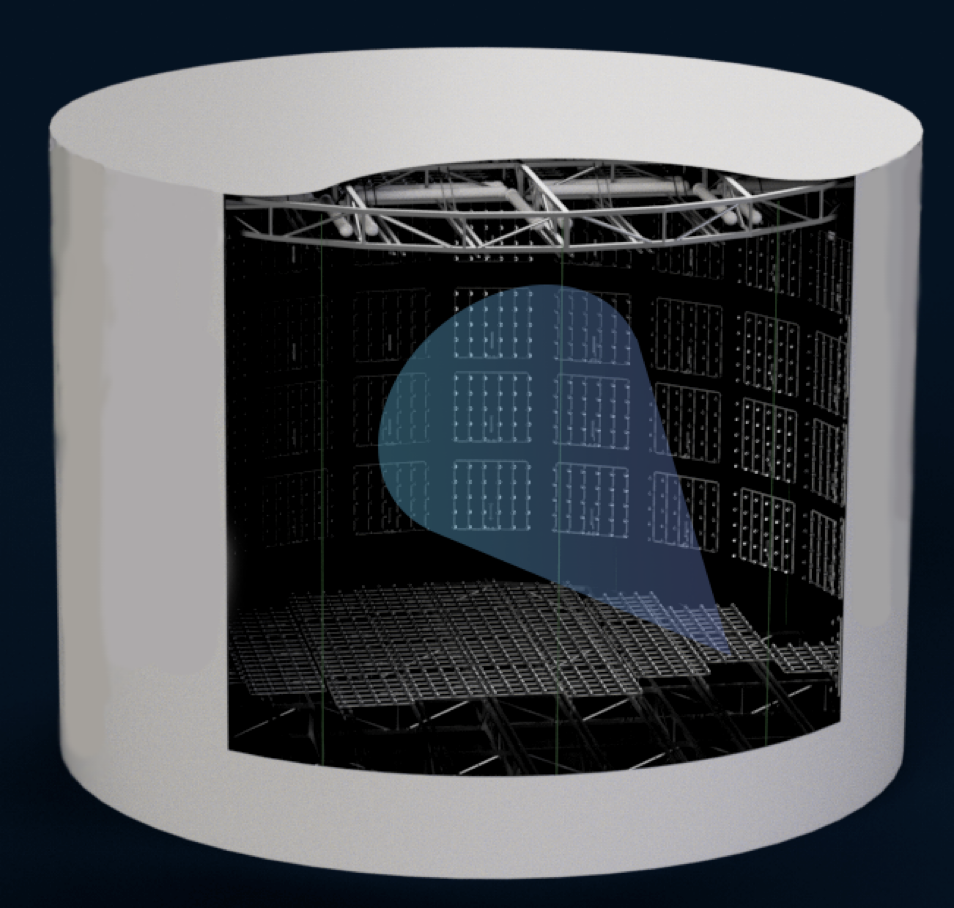}
    \caption[Graphical rendering of the \chipsfive detector]
    {Graphical rendering of the fully deployed and submerged \chipsfive detector module with a section of the liner cutaway. The bottom endcap and wall instrumentation is visible, as well as the top endcap structure and floatation. The faint green lines indicate the cables holding the endcaps together.}
    \label{fig:chips_render}
\end{figure}

The \chips detector module is a cylindrical water Cherenkov detector submerged in a deep body of water on the Earth's surface, such as a lake, reservoir, or as in the first prototype, a flooded mine pit. A graphical rendering of which is shown in Figure~\ref{fig:chips_render}. The water above the sunken detector provides a modest overburden from cosmic rays, while the surrounding water supports a lightweight structure. By removing the need for underground excavation and expensive structural support, the cost of construction can be dramatically reduced. On the other hand such bodies of water are not available everywhere, and some compromise between baseline distance and off-axis angle must be incorporated to maximize the sensitivity.

Additionally, the common practice of using majority bespoke components is replaced by using modern commercially available components wherever possible. The number of expensive elements, such as photomultiplier tubes, are also reduced by only considering multi-\GeV \ accelerator beam neutrino events, such that full high-density detector instrumentation is not required.

Furthermore, \chips detectors are not only designed to be cheap but practical. Easy to build, quick to deploy, and upgradable once operational, multiple detector modules can be flexibly combined depending on available resources and funding. Compared to DUNE and Hyper-Kamiokande, both of which require a large upfront budget and many years to construct, cheap \chips detector modules can be deployed as needed in under two years by a relatively small team.

\subsection{Modern Machine Learning for CHIPS}
\label{sec:intro_previous}
Alongside work to realise the concept outlined above, much effort has been made to explore alternative water Cherenkov neutrino event reconstruction and classification techniques for \chips style detectors. This work is motivated by the need to maximise the achievable physics performance from a `cheap as chips' detector.

Over the last few years, neutrino experiments have adopted modern machine learning techniques for a range of event analysis tasks~\cite{psihas2020}. In 2016 the \nova experiment applied a Convolutional Neural Network (CNN)~\cite{fukushima1982}, a type of deep learning~\cite{goodfellow2016} neural network, to the task of classifying the interaction type of events within their sampling calorimeter detector~\cite{aurisano2016}. Two views of raw detector data were used as input to train a CNN network~\cite{szegedy2015}. Further \nova iterations have since been applied to both the classification of individual energy deposit clusters~\cite{psihas2019} and $\nu_{e}$ and $e^{-}$ energy reconstruction~\cite{baldi2019}.

Another example of modern machine learning techniques in neutrino physics is the usage of deep neural networks by the IceCube collaboration, who, similar to \chips make use of a Cherenkov detector~\cite{kronmueller2019}. They have employed a classifier to distinguish between the charged current (CC) interactions of cascade causing neutrinos and the tracks of Cherenkov radiation-emitting muons as they traverse the ice.

CNNs have also been applied to liquid argon time-projection chambers. The MicroBooNE experiment~\cite{acciarri2017_ref} has shown that in addition to classification tasks and particle identification~\cite{abratenko2021}, the spatial localisation of single particles within events is also possible~\cite{acciarri2017}. Furthermore, the DUNE collaboration has designed a network to output both the interaction class and counts of different particle types within an event~\cite{collaboration2020, abi2020}, as well as kinematic energy estimation~\cite{liu2020deep}. This approach is called \emph{multi-task} learning and is discussed in detail in Section~\ref{sec:outputs}. The DUNE collaboration has also explored raw data denoising, the first step of reconstruction, in ProtoDUNE~\cite{Rossi2022}, with Graph Neural Networks also being investigated as an alternative to CNNs.

Submanifold sparse convolutional networks (SSCNs) have also been suggested to account for the poor scalability of naive applications of CNNs in large scale liquid argon time-projection chambers such as DUNE. The SSCNs can be used for image classification and object detection in such detectors and can have over 95\% pixel clustering efficiency and purity for Michel electrons~\cite{domine2020}. There are also investigations of the efficiencies of different CNN based algorithms such as Residual Networks and Inception Networks algorithms reconstructing shower energies in a LArTPC~\cite{carloni2022}.

Applications to water Cherenkov detectors have also been made by both the Daya Bay experiment~\cite{racah2016} and the KM3NeT/ORCA collaboration~\cite{aiello2020}. The output from a water Cherenkov detector, such as those envisioned by the \chips concept, is a simple \emph{image} of each event where two pieces of information are known for each photomultiplier tube (PMT): the number of collected photoelectrons and the associated hit times. Therefore, it is natural to use CNNs primarily developed for image-based computer vision tasks for \chips event characterisation.

This work outlines an event analysis methodology for \chips following this natural progression. Three forms of a CNN have been developed: One for cosmic muon rejection, one for beam event classification, and one for neutrino energy estimation, all using only a slightly modified version of the raw detector event as input.

For completeness, we discuss the detector simulation and then detail how CNNs are used for \chips. We then present a comprehensive analysis of the new methodology's performance. For evaluation purposes, only the implementation as applied to the \chipsfive prototype detector module is considered in this work. 

\section{Detector Simulation}
\label{sec:simulation}
The detector simulation used throughout this work~\cite{chipssim2020} builds upon the WCSim water Cherenkov simulation package~\cite{wcsim2020} which employs the Geant4 simulation framework~\cite{agostinelli2003, allison2006, allison2016}. Developed initially to simulate possible water Cherenkov detectors in the Long Baseline Neutrino Facility beam~\cite{acciarri2016}, WCSim is now used more widely in the field. 

The simulation builds an n-sided, regular polygonal prism consisting of two endcaps and a barrel, filled with water and lined with a low reflectivity \emph{blacksheet}. The geometry is separated into \emph{regions} within both the barrel and endcaps. Each region is filled with a unique base unit of geometry known as the \emph{unit cell}.

The unit cell defines a pattern of any number of PMTs, including their relative positions and in which direction they face. The final geometry is built by tiling the defined regions with their respective unit cell scaled to match the required regional photocathode coverage. Although exact detector PMT positions are not explicitly defined using this procedure, A given configuration will deterministically generate the same geometry. 

In this work, the \chipsfive geometry is generated with 28 sides and regions matching the photocathode coverage of the \chipsfive detector design. A conservative photon attenuation length of \SI{50}{\text{m}} at \SI{405}{\text{nm}} is used alongside negligible scattering~\cite{campbell2020}, with the PMT glass reflectivity set to 24\%~\cite{hamamatsu_handbook}, and the blacksheet reflectivity kept at the WCSim default of 4\%. Note that veto PMTs are ignored for simplicity.

Using the expected \numi flux at \chipsfive, the \textsc{Genie} neutrino event generator (version 3.0.6)~\cite{andreopoulos2009, andreopoulos2015} is used to generate beam neutrino events using default neutrino cross-sections on water.  All channels of neutrino interaction are considered and can be broken down into five main categories:
\begin{itemize}
    \item \textbf{Quasi-Elastic scattering (QEL):} The dominant channel for energies below 1 GeV, involving the neutrino scattering off the entire nucleon.
    \item \textbf{Meson Exchange Current (MEC):} Additional contribution channel for energies below 1 GeV, involving two nucleons and producing two protons in the final state.
    \item \textbf{Resonant pion production (Res)} The dominant channel between 1 and 2 GeV, involving the neutrino exciting the nucleon into a resonant state.
    \item \textbf{Coherent pion production (Coh)} A mechanism where the neutrino scatters coherently from the entire nucleus.
    \item \textbf{Deep Inelastic Scattering (DIS)} The dominant channel for neutrino energies above 3 GeV with the additional energy allowing for the neutrino to resolve the individual quark content of the nucleon.
\end{itemize}
The Cosmic-Ray Shower Library (CRY)~\cite{hagmann2012_1, hagmann2012_2} is used for cosmic ray event generation, assuming a \chipsfive overburden of \SI{50}{\text{m}} and a \SI{2.2}{\MeV/\text{cm}^{2}} muon energy loss in water as suggested by~\cite{klimushin2001}.

Beam event vertices are randomly placed within the inner detector volume, while cosmic vertices are set at \SI{1}{\text{m}} above the detector volume. WCSim then simulates the passage of all particles through the detector materials, with interactions, decays, and Cherenkov emission all considered. Whenever a photon is calculated to have hit the photocathode of a PMT, an angular dependent acceptance efficiency is applied to see if it is recorded~\cite{hamamatsu_handbook}. If accepted, all hits within \SI{200}{\text{ns}} windows are grouped to form a single recorded hit, with a smeared first hit time used as the recorded time. The standard WCSim methodology is used to determine the total output charge of the hit, given the number of incident photons. This procedure involves a single photoelectron charge distribution being repeatedly probed for each photon before the combined sum is returned which corresponds to the sum of the simulated responses for each photon~\cite{tutorial2020}. The simulation output, used as the input for the rest of this work, is a collection of hits for each neutrino interaction event, each with an associated number of photoelectrons and hit time.

\section{Convolutional Neural Networks for CHIPS}
\label{sec:cnn}
For the majority of HEP experiments, event analysis entails the separation of signal from background, the identification of particle types, the determination of spatial properties, and the estimation of energies. The same is true for \chips detectors, with the primary aims being the selection of appeared CC $\nu_{e}$ signal beam events from a sizeable background (beam and cosmic) and the estimation of associated neutrino energies.

For this purpose, the \chips project has so far relied on a likelihood-based reconstruction algorithm and a simple classification neural network driven by hand-engineered features~\cite{chipsreco2020}. Both suffer from only considering what has been implemented in software and consequently what features are explicitly extracted and then modelled from the data. This restriction makes them prone to ignoring the wide range of edge cases not contained within the bulk of neutrino events and unable to use all the underlying informative features of the data.

The methods outlined in this section present a replacement event analysis methodology for \chips. As mentioned previously, the \emph{image} like nature of the raw detector output is a natural fit for the standard grid-like input to a CNN. Notably, using a CNN on the raw detector event removes the requirement to build hand-engineered features as the CNN learns to extract the most powerful features within the input data itself~\cite{alzubaidi2021review}. Three forms of a baseline CNN have been developed to achieve the primary aims outlined above:
\begin{itemize}
    \item The \textbf{cosmic rejection} form aims to prevent the vast cosmic muon background from contaminating the final selected sample of beam \linebreak events. Therefore, the primary task is a simple binary classification between beam-like and cosmic-like events.
    \item The \textbf{beam classification} form aims to separate beam events by their neutrino and interaction type to primarily select a pure and efficient sample of appeared CC $\nu_{e}$ events, but also a sample of survived CC $\nu_{\mu}$ events. Therefore, the principal task is a categorical classification between CC $\nu_{e}$, CC $\nu_{\mu}$, and background Neutral Current (NC) events.
    \item The \textbf{neutrino energy estimation} form aims to accurately estimate the energy of the signal event $\nu_{e}$ and $\nu_{\mu}$. Therefore, the primary task is a regression on the interactive neutrino energy.
\end{itemize}

A Python-based software package named \emph{chipsnet}~\cite{chipsnet2020} has been built for this purpose. By using the high-level \emph{Application Programming Interfaces} (APIs) provided by the Tensorflow framework (version 2.3.0)~\cite{tf2015}, a complete pipeline including data preparation, network training, and performance evaluation has been implemented. We separate our methodology into four sections: the inputs used, the CNN architecture, the CNN outputs, and finally the training procedure. For each, we detail the differences between the three network forms.

\subsection{Inputs}
\label{sec:inputs}
The primary difficulty in applying CNNs to \chips (or any cylindrical detector) is determining how to map an event captured on a cylindrical surface to a two-dimensional grid. This must be done in such a way that the underlying Cherenkov emission topology is not distorted, which could inhibit network learning. This work builds upon the ideas outlined in reference~\cite{theodore2016}. Simply put, an event is mapped onto a two-dimensional grid as though it is viewed from its estimated interaction vertex position. The primary motivation behind this is to remove any detector shape effects and focus on the underlying event topology and Cherenkov emission profiles.

To estimate the interaction vertex position, the PMT hits of an event are firstly sliced in both space and time. Gaps in the time ordering of hits are used to separate the event into time slices. Each of these slices then undergoes basic clustering to remove outlying hits and ensure only the dominant collections of hits are considered. Each cleaned slice is then run through a simple geometric vertex finding algorithm to estimate the interaction vertex position.

A circular Hough transform algorithm, traditionally used for water\linebreak Cherenkov ring finding, is then applied~\cite{illingworth1988}. As output, the voting-based transformation produces a space within which rings of PMT hits exist as peaks. The interaction vertex position is further refined using the Hough peaks in this space.

Using $\theta$ and $\phi$ components calculated as viewed from the estimated interaction vertex position facing along the beam axis, hit PMTs are mapped onto a $64 \times 64$ grid. This procedure is used to generate two event \emph{maps}. Firstly, a \emph{hit-charge} map where each grid bin is given by the total collected photoelectrons from all PMTs mapped to that bin. Secondly, a \emph{hit-time} map where each grid bin is given by the first hit time (in nanoseconds) across all PMTs mapped to that bin. Each hit-time map is further corrected so that the first hit time across all bins lies at zero.

By design, the Hough transform uses the estimated interaction vertex position to generate the transform space. Therefore, by re-binning the transform space to a $64 \times 64$ grid, a third \emph{hough-height} map is generated for each event. This event map provides a complementary but different representation of the event where Cherenkov rings are instead represented as peaks, allowing for additional discriminating features to be learnt.

\begin{figure}
    \centering
    \includegraphics[width=\textwidth]{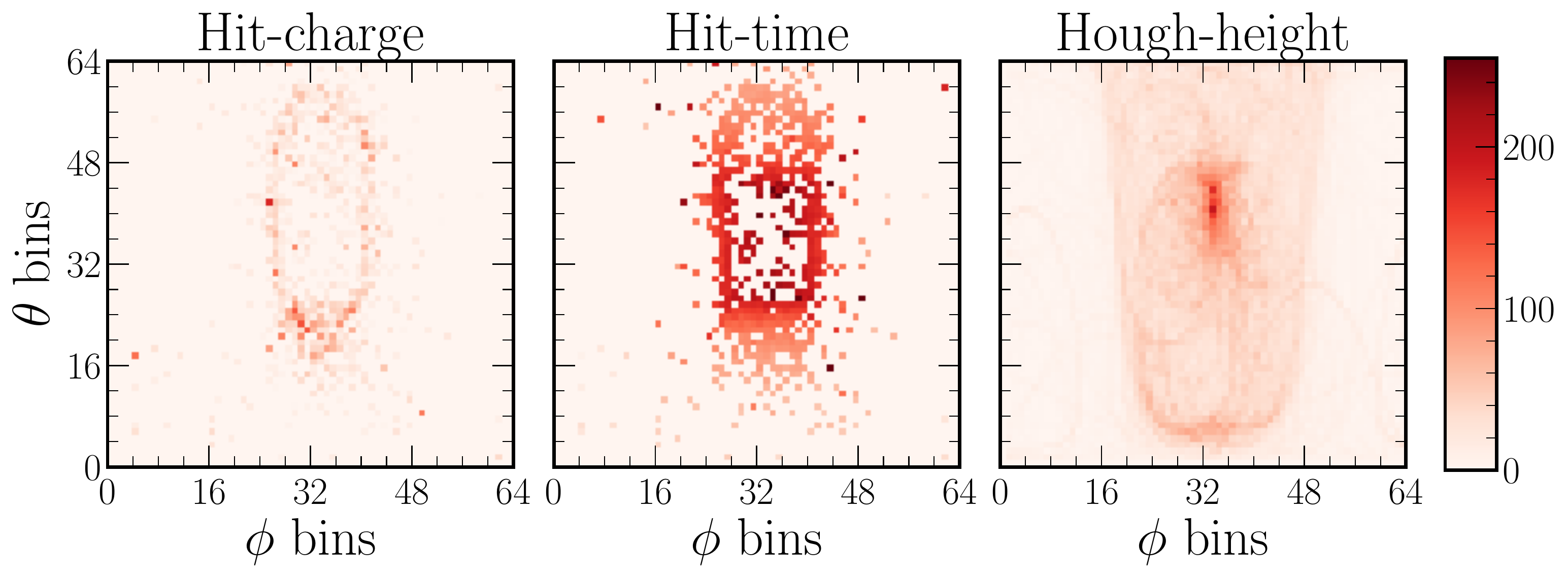}
    \caption{Three event map representations of a CC resonant $\nu_{e}$ event expected within \chipsfive. Initiated by a $\nu_{e}$ of energy \SI{3.3}{\GeV} the final state particles above the Cherenkov threshold include an $e^{-}$ of energy \SI{2.8}{\GeV} and a \SI{0.3}{\GeV} $\pi^{0}$.}
    \label{fig:example_event}
\end{figure}

All three event maps: hit-charge, hit-time, and hough-height, are down-sampled using 8-bit encoding by converting each 32-bit float value to an integer between 0 and 255. Encoding significantly reduces storage requirements and dramatically increases the speed with which data can be loaded during training (which is the primary training bottleneck). For each map type, a range over which to encode from zero up to a \emph{cap-point} is chosen to minimise the number of bin values that are capped at the maximum encoded value of 255. The generated event maps for an example CC $\nu_{e}$ event are shown in Figure~\ref{fig:example_event}. Each network form (cosmic rejection, beam classification, and energy estimation) uses the same event maps as input.

\subsection{Architecture}
\label{sec:architecture}
An illustrative diagram of the baseline chipsnet CNN architecture is shown in Figure~\ref{fig:chipsnet}. Based on the Visual Geometry Group (VGG) network~\cite{simonyan2014} there are a few key differences from the literature defined network:

\begin{figure}
    \centering
    \includegraphics[width=0.7\textwidth]{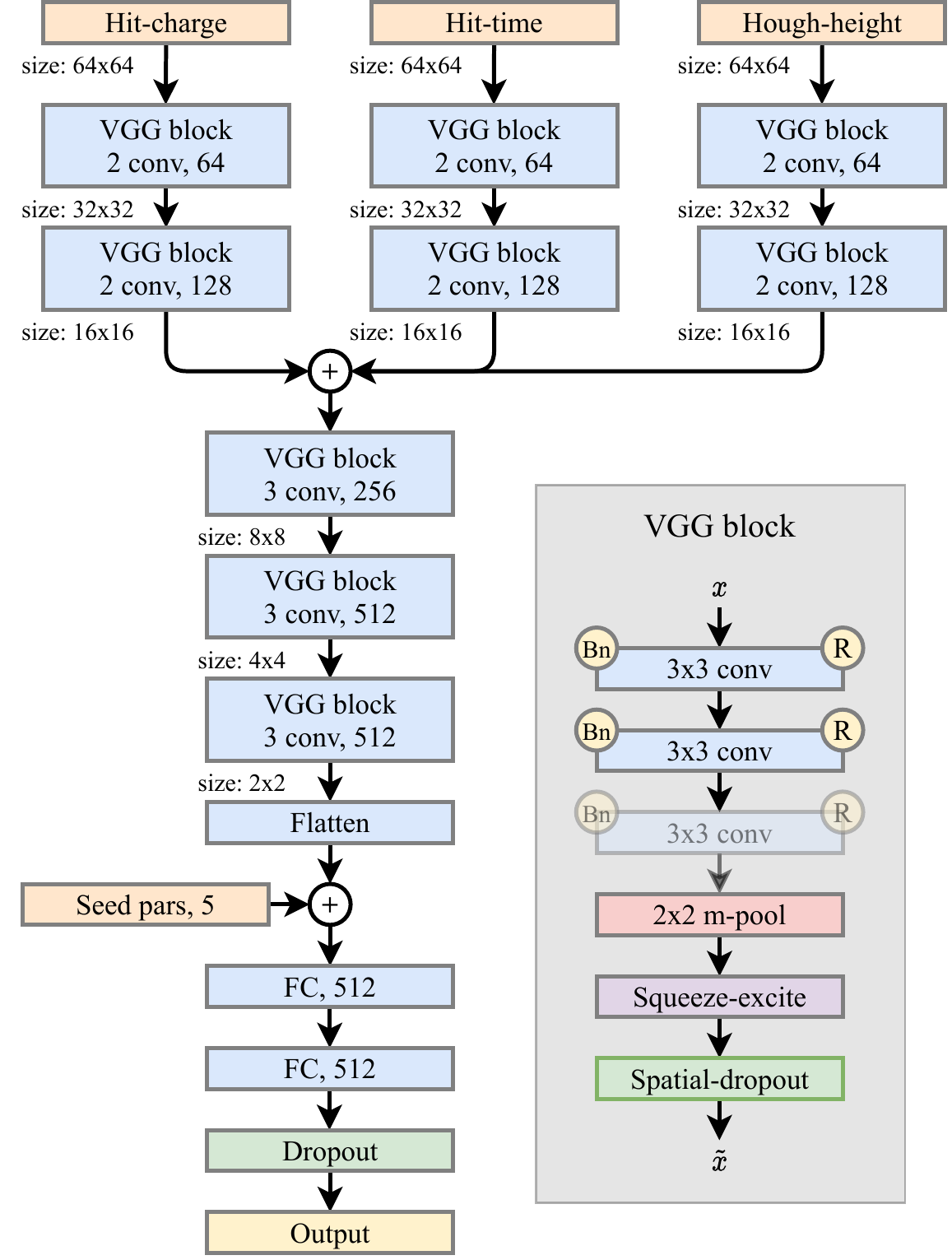}
    \caption{Illustrative diagram of the baseline chipsnet architecture. The three input event maps are separately passed through two VGG blocks each before their outputs are combined and passed through a further three VGG blocks together. The flattened VGG blocks outputs are then concatenated with five seed parameters (seed pars) and passed through two fully-connected (FC) layers of 512 neurons each before the output layer. Both the number of convolutional units (1st value) and kernels (2nd value) is shown for each block. The detailed VGG block structure is shown within the grey box. The circular yellow $R$ and $Bn$ indicate the use of the ReLU activation function and batch normalisation, respectively.}
    \label{fig:chipsnet}
\end{figure}

\begin{itemize}
    \item Each of the three event maps: hit-charge, hit-time, and hough-height, are initially fed into three separate branches. Each branch contains two VGG blocks with two convolutional (conv) layers each (four convolutional layers in total). The outputs from each branch are merged using a concatenation layer before being fed to the rest of the network. This configuration allows for event map specific features to be learnt independently before combined features are learnt by the rest of the network.
    \item Batch normalisation~\cite{ioffe2015} is included before the activation (ReLU) function for every convolutional layer.
    \item Squeeze-and-excitation units~\cite{hu2018}, are included after the max-pooling operation in all VGG blocks. These units introduce extra parameters to model the interdependencies between output feature maps, allowing the network to learn how to weight each feature map effectively.
    \item Dropout is included at the end of each VGG block as well as after the final fully-connected layer. Instead of dropping individual kernel elements, the dropout within the VGG blocks drops entire kernels at each training iteration, this is commonly called \emph{two-dimensional spatial dropout}. The dropout after the fully-connected layers is standard, in that it drops out individual fully-connected neurons.
    \item Five parameters from the estimation of the interaction vertex position (seed pars) are concatenated with the flattened layer before the fully-connected layers. Included are the three components of the estimated interaction vertex position ($s_{x},s_{y}, s_{z}$), and the two components of the estimated track direction ($s_{\theta},s_{\phi}$). These parameters provide the network with spatial context as to where the input event maps have been generated in the detector and the dominant direction of PMT activity.
\end{itemize}

The chipsnet baseline architecture is implemented using the Keras API built into Tensorflow~\cite{chollet2015}. Each network form (cosmic rejection, beam classification, and energy estimation) uses the same baseline model architecture.

\subsection{Outputs}
\label{sec:outputs}
Many CNN applications are found to benefit from learning multiple tasks simultaneously. This is believed to be because training with multiple tasks tends to return a network with an improved generalised representation of the inputs, with features learnt for one task improving the performance of another. Additionally, multiple tasks work to prevent any one output from overfitting. Commonly named \emph{multi-task} learning, this methodology is used within this work.

To train a network with multiple tasks (outputs), a loss function $E_{tot}$, must be defined to combine the individual loss functions for each task $E_{i}$. The simplest way to do this is via a linear weighted sum, such that
\begin{equation}
    E_{tot} = \sum_{i=1}^{i=N}w_{i}E_{i},
    \label{eq:multi_simple}
\end{equation}
where $N$ is the number of tasks and $w_{i}$ are the associated weights. In this work this is referred to as the \emph{simple} multi-task loss.

The final network performance can strongly depend on the relative weighting between loss functions, especially when the values returned by each differ by many order of magnitude (common when combining regression and classification tasks). Therefore, finding the optimal $w_{i}$ weights can be both difficult and time-consuming. Another approach outlined in reference~\cite{kendall2018} remedies this problem by learning the optimal weighting between loss functions. This is done by introducing an additional trainable parameter $\sigma_{i}$, for each loss function, such that
\begin{equation}
    E_{tot}= \sum_{i=1}^{i=N}\frac{1}{2\sigma_{i}^2}E_{i}+ \log\sigma_{i}.
    \label{eq:multi_learnt}
\end{equation}
In this work we refer to this as the \emph{learnt} multi-task loss.

The specific number and nature of outputs for the specific network forms are detailed below. Although physically motivated to some degree, the exact set of tasks and the loss combination technique used is mainly driven by extensive trial-and-error.

\subsubsection{Outputs - Cosmic Rejection}
Alongside the primary task of classification between beam-like and cosmic-like events, training the network to also separate events where the primary charged lepton escapes the detector volume or not, is found to improve cosmic rejection performance. As a large proportion of cosmic muons are relatively high in energy and, therefore, escape the detector in this fashion, there is motivation as to why this additional task is helpful. Hence the two outputs for this network form are:
\begin{enumerate}
    \item \textbf{Cosmic score (1 classification neuron):} Returns a score between zero and one corresponding to whether the event is beam or cosmic like.
    \item \textbf{Escapes score (1 classification neuron):} Returns a score between zero and one corresponding to whether the charged lepton in an event is contained or escapes the
    detector. NC beam events without a charged lepton are masked (do not contribute to the loss) during training for this output.
\end{enumerate}
Both outputs are trained using a binary cross-entropy loss function and combined using the \emph{simple} multi-task loss in Equation~\ref{eq:multi_simple}, each with a weight of $1$.

\subsubsection{Outputs - Beam Classification}
Alongside the primary task of classification between CC $\nu_{e}$ , CC $\nu_{\mu}$, and background NC events, training the network on additional classification and particle counting tasks is found to improve performance. Note that the particle counting tasks are not used in this work for anything but to increase the primary classification performance. However, future work could exploit any ability to separate exclusive final states, deduced from these particle counts, to improve energy resolution and systematic errors. Hence the nine outputs for this network form are:
\begin{enumerate}
    \item \textbf{Combined category (3 classification neurons):} Returns a classification probability score between zero and one for each of CC $\nu_{e}$, CC $\nu_{\mu}$, and NC (summing to one).
    \item \textbf{CC category (6 classification neurons):} Returns a classification probability score between zero and one for each of CC, Quasi-Elastic (QEL),  Deep Inelastic Scattering (DIS), Resonant (Res), \linebreak Coherent (Coh), Meson Exchange Current (MEC), and CC-other (summing to one). NC events are masked (do not contribute to the loss) during training for this output.
    \item \textbf{NC category (4 classification neurons):} Returns a classification probability score between zero and one for each of NC-Res, NC-DIS, NC-Coh, and NC-other (summing to one). CC events are masked (do not contribute to the loss) during training for this output.
    \item \textbf{Electron count (4 classification neurons):} Returns a classification probability score between zero and one for each of 0, 1, 2, and 3+ electrons in the final state (summing to one).
    \item \textbf{Muon count (4 classification neurons):} Returns a classification probability score between zero and one for each of 0, 1, 2, and 3+ muons in the final state (summing to one).
    \item \textbf{Proton count (4 classification neurons):} Returns a classification probability score between zero and one for each of 0, 1, 2, and 3+ protons in the final state (summing to one).
    \item \textbf{$\pi^{\pm}$ count (4 classification neurons):} Returns a classification probability score between zero and one for each of 0, 1, 2, and 3+ charged pions in the final state (summing to one).
    \item \textbf{$\pi^{0}$ count (4 classification neurons):} Returns a classification probability score between zero and one for each of 0, 1, 2, and 3+ neutral pions in the final state (summing to one).
    \item \textbf{Photon count (4 classification neurons):} Returns a classification probability score between zero and one for each of 0, 1, 2, and 3+ photons in the final state (summing to one).
\end{enumerate}
All outputs are trained using a categorical cross-entropy loss function and combined using the \emph{simple} multi-task loss in Equation~\ref{eq:multi_simple}, each with a weight of $1$.

\subsubsection{Outputs - Energy estimation}
Alongside the primary task of estimating the neutrino energy, training the network to additionally estimate the primary charged lepton energy and the interaction vertex position and time are found to improve performance. Although this improvement is relatively small for neutrino energy estimation, it dramatically
improves primary charged lepton energy estimation. With two energy tasks, the network is encouraged to learn how the primary charged lepton and neutrino energies are related. As the interaction vertex position within the detector and hence distance from the wall can impact the number of deposited photoelectrons, there is motivation as to why this additional task is also helpful. Hence the six outputs for this network form are:
\begin{enumerate}
    \item \textbf{Neutrino energy (1 regression neuron):} Returns the estimated neutrino energy.
    \item \textbf{Charged lepton energy (1 regression neuron):} Returns the estimated charged lepton energy.
    \item \textbf{Interaction vertex x-position (1 regression neuron):} Returns the estimated interaction vertex x-position.
    \item \textbf{Interaction vertex y-position (1 regression neuron):} Returns the estimated interaction vertex y-position.
    \item \textbf{Interaction vertex z-position (1 regression neuron):} Returns the estimated interaction vertex z-position.
    \item \textbf{Interaction time (1 regression neuron):} Returns the estimated interaction time.
\end{enumerate}
All outputs are trained using a mean-squared error loss function and combined using the \emph{learnt} multi-task loss in Equation~\ref{eq:multi_learnt} which includes the additional trainable parameters.

\subsection{Training}
\label{sec:training}
All networks are trained on an 18 core CPU (36 thread) machine equipped with four NVIDIA GeForce RTX 2080 graphics processing units (GPUs). The Tensorflow dataset API is used to create an efficient input data pipeline where data is loaded on-the-fly at training time. This procedure ensures all CPU threads are utilised loading, decoding, and preprocessing data for the primary GPU based network calculations before being needed, maximising computational efficiency.

During preprocessing, all 8-bit input event map values are converted to 32-bit float values
bounded between zero and one. A random factor scaling is applied to each map bin, generated from a normal distribution centred on one with a standard deviation of $\sigma_{r}$. By fluctuating the bin values the network is forced to focus less on the absolute bin values and more on the underlying event topology. This process provides valuable regularisation to reduce overfitting and makes the trained networks robust to small changes within the input.

A minibatch training strategy of minibatch size of $n_{b}$, using the Adam optimiser~\cite{kingma2014} ($\beta_{1}=0.9$, $\beta_{2}=0.999$, and $\epsilon = 10e-7$) is used. The exact training sample size and composition for each specific network are given in below, but for all network forms a 95\% training to 5\% validation data split is employed across the full training sample. The learning rate for each epoch $\eta_{e}$, is set to decrease throughout training according to
\begin{equation}
    \eta_{e}=\frac{\eta_{0}}{1+c_{d}(e-1)},
\end{equation}
where $\eta_{0}$ is the initial learning rate, $e$ is the epoch number (starting at one), and $c_{d}$ is the learning rate decay coefficient. Early stopping is also used to stop training once the network form specific performance metric has plateaued. 

When training each network, there is a list of tunable hyperparameters, all of which are optimised using the SHERPA hyperparameter tuning framework~\cite{hertel2020}. To maximise performance, SHERPA uses a random search algorithm to select random configurations of hyperparameters which are then tested by training the network for five epochs on the available training data. Each configuration's performance is assessed by using a metric detailed for each network below. The search space is confined to a specific range or selection of choices for each hyperparameter, with:
\begin{itemize}
    \item the \textbf{initial learning rate $\eta_{0}$}, in a range from $0.00005$ to $0.001$;
    \item the \textbf{learning rate decay coefficient $c_{d}$}, in a range from $0.2$ to $0.8$;
    \item the \textbf{dropout probability $p_{d}$}, in a range from $0.0$ to $0.5$;
    \item the \textbf{random scaling size $\sigma_{r}$}, in a range from $0.0$ to $0.1$; and
    \item the \textbf{minibatch size $n_{b}$}, choosing from either $32$, $64$, $128$, or $256$;
\end{itemize}

\subsubsection{Training - Cosmic Rejection}
\begin{figure}
    \centering
    \includegraphics[width=0.7\textwidth]{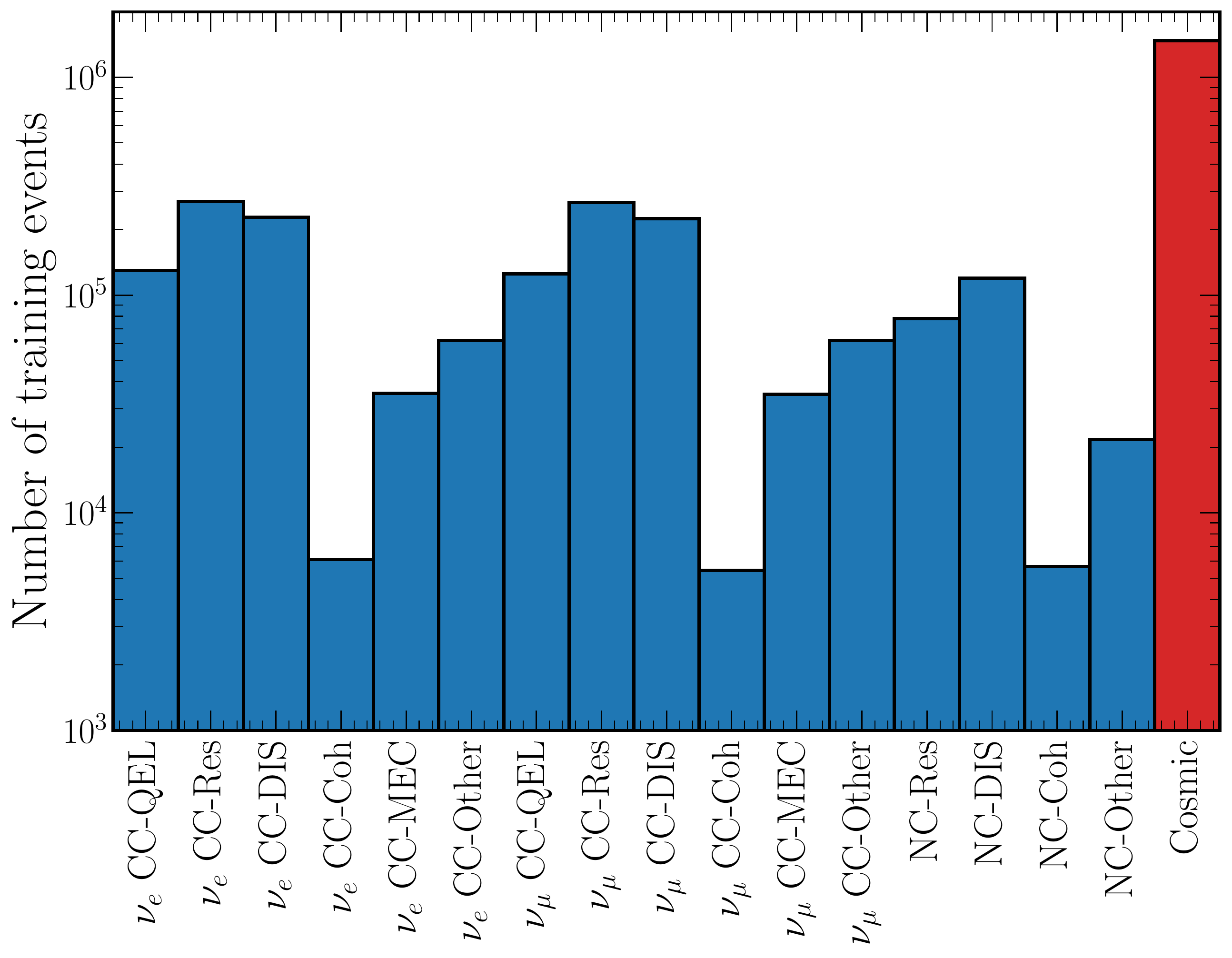}
    \caption{The number of training events per category for the cosmic rejection and beam classification network forms. In the cosmic rejection case, all beam event interaction types (shown in blue) are classed as beam events against the cosmic events (shown in red). In the beam classification case, only those events shown in blue are used.}
    \label{fig:training_sample}
\end{figure}

The cosmic rejection network form is trained on a sample of 3.15 million simulated events produced using the detector simulation and event generation methods outlined in Section~\ref{sec:simulation}. Roughly $1/3^{rd}$ are $\nu_{\mu}$ beam events, $1/3^{rd}$ $\nu_{e}$ beam events, and $1/3^{rd}$ cosmic muon events, the counts of which are shown in Figure~\ref{fig:training_sample}. All beam events (both $\nu_{\mu}$ and $\nu_{e}$) are generated using the expected unoscillated \chipsfive $\nu_{\mu}$ energy spectrum to closely mimic the dominant $\nu_{\mu}$ beam component and appeared $\nu_{e}$ signal.

The network is allowed to train for up to 30 epochs using the SHERPA optimised hyperparameters: $\eta_{0}=0.00005$, $c_{d}=0.7$, $p_{d}=0.1$, $\sigma_{r}=0.02$, and $n_{b}=128$. The \emph{cosmic score} accuracy metric, defined as the fraction of validation sample events that are correctly classified when a cut value of 0.5 on the \emph{cosmic score} output is used to determine the classification
of each event and is used for SHERPA optimisation and early stopping.

\subsubsection{Training - Beam Classification}
The beam classification network form is trained on a 1.67 million event subset of the training sample used for the cosmic rejection network form, excluding the cosmic muon events. The event counts are again shown in Figure~\ref{fig:training_sample}.

The network is allowed to train for up to 30 epochs using the SHERPA optimised hyperparameters: $\eta_{0}=0.0002$, $c_{d}=0.5$, $p_{d}=0.1$, $\sigma_{r}=0.02$, and $n_{b}=128$. The \emph{combined
category} accuracy metric, defined as the fraction of validation sample events that are correctly classified when the highest scoring \emph{combined category} output neuron is used to determine the classification of each event and is used for SHERPA optimisation and early stopping.

\subsubsection{Training - Energy estimation}
\label{sec:cnn_energy_training}
Accurate neutrino energy estimation is accomplished using multiple energy estimation networks trained on separate samples of $\nu_{e}$ and $\nu_{\mu}$ events across multiple CC interaction types. It is found that separation such as this results in greater performance compared to if a single energy estimation network form or even separate $\nu_{e}$ and $\nu_{\mu}$ network forms are trained. This is expected as a single set of network weights is unlikely to capture the specific topological features that contribute to the energy for all types of event.

Separate energy estimation network forms are trained for each of CC-QEL (and CC-MEC), CC-Res, and CC-DIS for both $\nu_{e}$ and $\nu_{\mu}$ events (6 in total) using 250000 corresponding simulated events each. All events (both $\nu_{\mu}$ and $\nu_{e}$) are produced using the detector simulation and event generation methods outlined in Section~\ref{sec:simulation}. Only events for which the primary charged lepton is fully contained within the detector volume are used for training. Note that CC-QEL and CC-MEC energy estimation is combined into a single network as both have incredibly similar final state topologies (a single charged lepton).

Each energy estimation network is allowed to train for up to 30 epochs using the SHERPA optimised hyperparameters: $\eta_{0}=0.0002$, $c_{d}=0.5$, $p_{d}=0.1$, $\sigma_{r}=0.0$, and $n_{b}=128$. The \emph{neutrino energy} mean absolute error metric, defined as the average difference between the true and estimated neutrino energies across all validation sample events is used for SHERPA optimisation and early stopping.

\section{Results}
\label{sec:results}
Before discussing the results, a hugely impactful advantage of the CNN approach must be highlighted. Although the time taken to train the CNNs in this work is approximately two days, once trained, the time required to calculate all network outputs (inference time) for a single event is on the order of \SI{2}{\text{ms}}. When combined with event seeding and event map generation, the total time taken to reconstruct and classify a raw event fully is less than \SI{0.1}{\text{seconds}}. Compared to the $\sim$\SI{15}{\text{minutes}} required for each event using the standard reconstruction and classification methods, the difference is stark.

Armed with this incredible speed, the time taken to fully process a large dataset containing millions of events becomes a matter of hours, compared to the many weeks typically required. This change has far-reaching implications for how physics analysis is conducted. By removing the processing bottleneck, larger datasets can be used without worry, new techniques can be tested quickly, and overall analysis turnover increased.

\subsection{Evaluation sample}
\label{sec:results_eval_sample}

An independent sample of events is used to evaluate the combined performance of the trained CNNs. The evaluation sample consists of 400000 beam and 350000 cosmic muon events produced in the same way as the training and validation events, using the detector simulation and event generation methods outlined in Section~\ref{sec:simulation}. The beam events include the expected $\nu_{\mu}$, $\bar{\nu}_{\mu}$, $\nu_{e}$ and $\bar{\nu}_{e}$ components as well as events generated to mimic the appeared $\nu_{e}$ component. Only the neutrino mode (forward horn current) of NuMI beam operation is considered here. During the evaluation, the beam's intrinsic neutrino and antineutrino components are considered the same for simplicity.

\begin{figure} 
    \centering
    \includegraphics[width=0.7\textwidth]{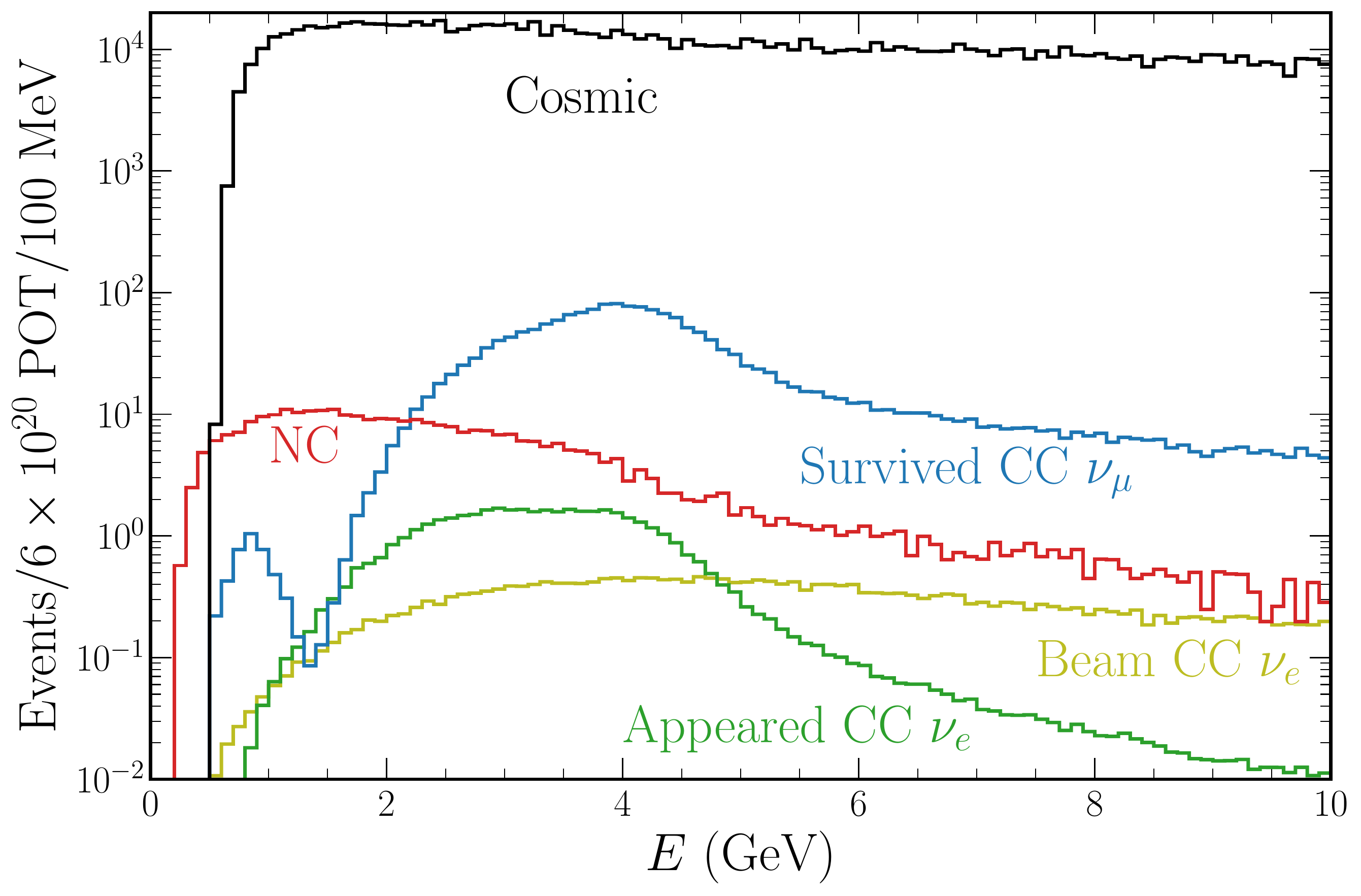}
    \caption{Weighted spectrum of events contained within the evaluation sample. The weighting is designed to mimic the expected event spectrum of the \chipsfive detector. Beam events are weighted by combining the expected unoscillated flux with cross-sections and standard oscillation probabilities, while cosmic events are weighted using the expected cosmic rate. Shown in blue, green, and olive are the surviving CC $\nu_{\mu}$, appearing CC $\nu_{e}$, and intrinsic beam CC $\nu_{e}$ spectra respectively, binned in terms of their neutrino energy. Shown in red is the NC event spectra, binned in terms of the energy of the hadronic component (excluding the outgoing neutrino energy) to represent more accurately the energy visible to the detector. Finally, shown in black is the cosmic muon event spectra binned in terms of the muon energy.}
    \label{fig:explore_osc_fluxes}
\end{figure}

All evaluation events are weighted to match the expected spectrum at the \chipsfive detector using the flux, cross-sections, and oscillation probabilities (derived from the NuFIT oscillation parameters~\cite{esteban2020}, assuming the normal hierarchy, and including matter effects). Additional weighting also scales the sample to match data taking in the NuMI beam for a single year, corresponding to $6\times 10^{20}$ protons on target (POT). Cosmic muon events are weighted according to a \SI{11.8}{\text{kHz}} expected \chipsfive rate over a full year. The final weighted spectrum of evaluation events is shown in Figure~\ref{fig:explore_osc_fluxes}.

\subsection{Preselection cuts}
\label{sec:results_cuts}
\begin{figure} 
    \centering
    \includegraphics[width=\textwidth]{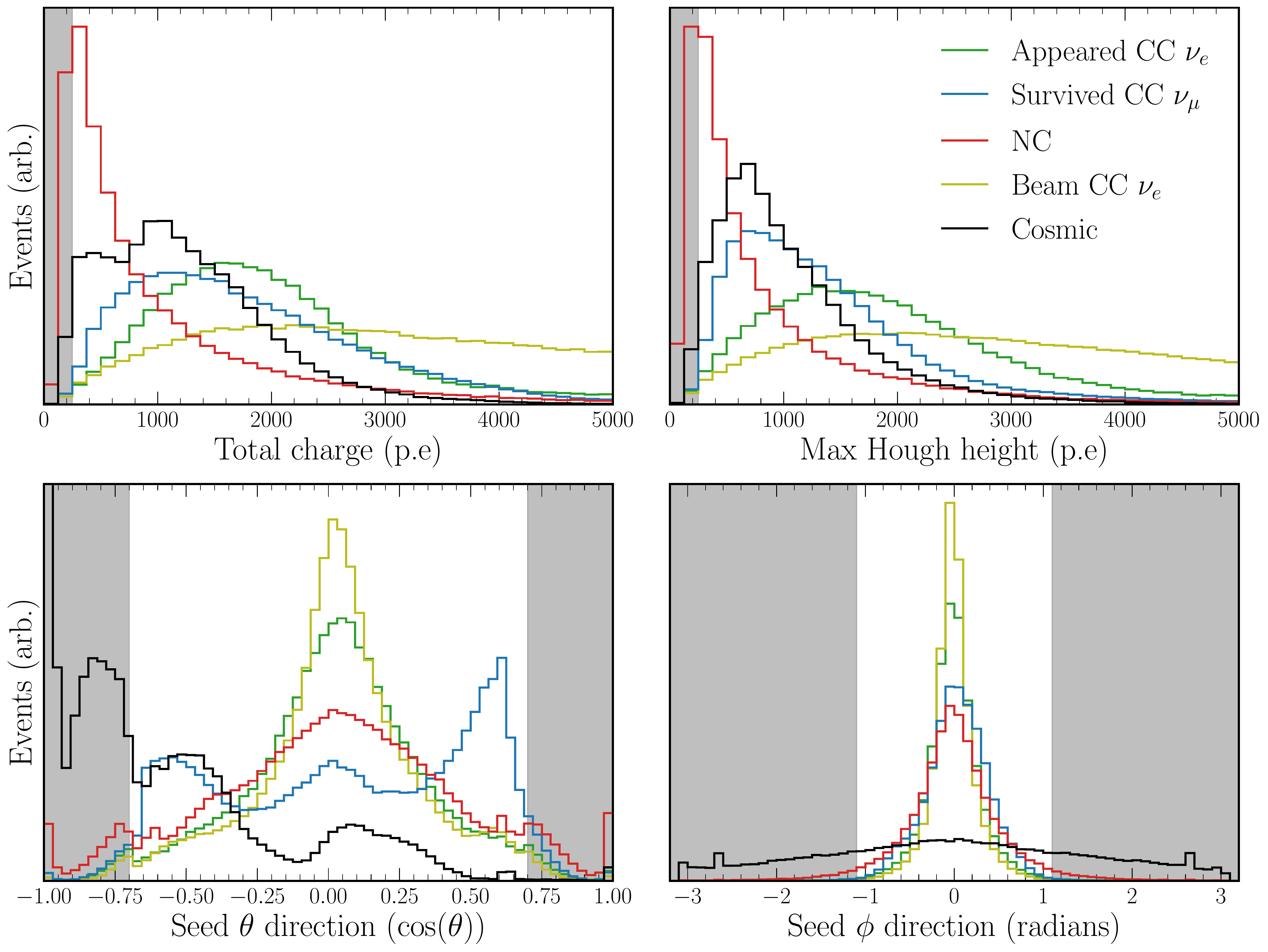}
    \caption{The four preselection cuts and their effect on the different event categories. The grey regions indicate rejected events.}
    \label{fig:explore_simple_cuts}
\end{figure}

Separate from the CNN driven work, a simple preselection is applied to all evaluation sample events. Designed to reject cosmic and NC events while keeping the selection efficiency of CC beam events high, the preselection consists of four simple cuts, shown in Figure~\ref{fig:explore_simple_cuts}. Firstly, the total number of collected photoelectrons (charge) across all PMTs in the event must be greater than 250. Secondly, the maximum Hough transform space height must be greater than 250 photoelectrons\footnote{This is dependent on the particular 64 by 64 binning used for the Hough transform.}. Thirdly, the seeding procedure $\cos(\theta)$ direction must be between $\pm$0.7. Finally, the seeding procedure $\phi$ direction must be between $\pm$1.1 radians. The first two cuts reject low energy events, typically NC, while the last two reject events whose activity is not along the beamline, typically cosmic.

\subsection{Cosmic rejection and containment} 
\label{sec:results_cosmic} 

\begin{figure} 
    \centering
    \includegraphics[width=0.7\textwidth]{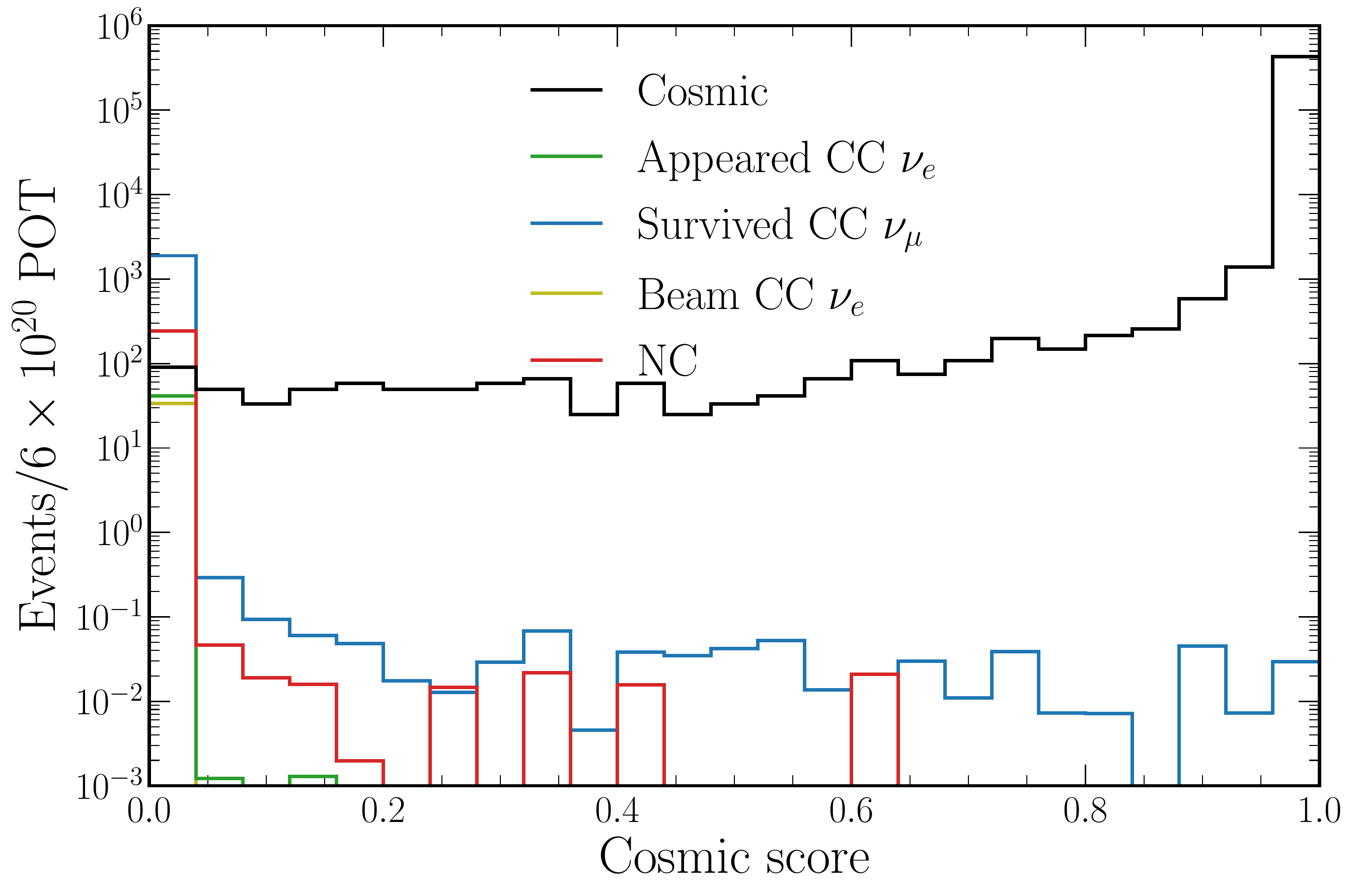}
    \caption{Distribution of \emph{cosmic score} output values from the trained cosmic rejection network for the different event categories. A score close to one signifies a cosmic-like event, while a score close to zero corresponds to a beam-like event. Only preselected events are shown to better highlight the events this output aims to classify.}
    \label{fig:cosmic_outputs}
\end{figure}

The \emph{cosmic score} output from the trained cosmic rejection network form shows an excellent separation between beam (output close to zero) and cosmic (output close to one) events, as can be seen in Figure~\ref{fig:cosmic_outputs}. Notably, $99.2\%$ of beam events are associated with a score $<0.0001$. Given this fact, a \emph{cosmic score} of below $0.0001$ is chosen to select beam-like events. Out of the total $350000$ cosmic events in the evaluation sample, all are rejected by this cut alongside preselection.

It is crucial for accurate neutrino energy estimation that the activity of an event is fully contained within the volume of the detector. If charged event particles instead leave the detector and emit Cherenkov radiation not captured by PMTs, estimating the resulting missing energy and, hence, neutrino energy can be incredibly difficult. Within the \chipsfive detector, this is particularly important for long track CC $\nu_{\mu}$ events for which only 44\% of the primary charged muons are fully contained within the detector volume.

\begin{figure} 
    \centering
    \includegraphics[width=0.7\textwidth]{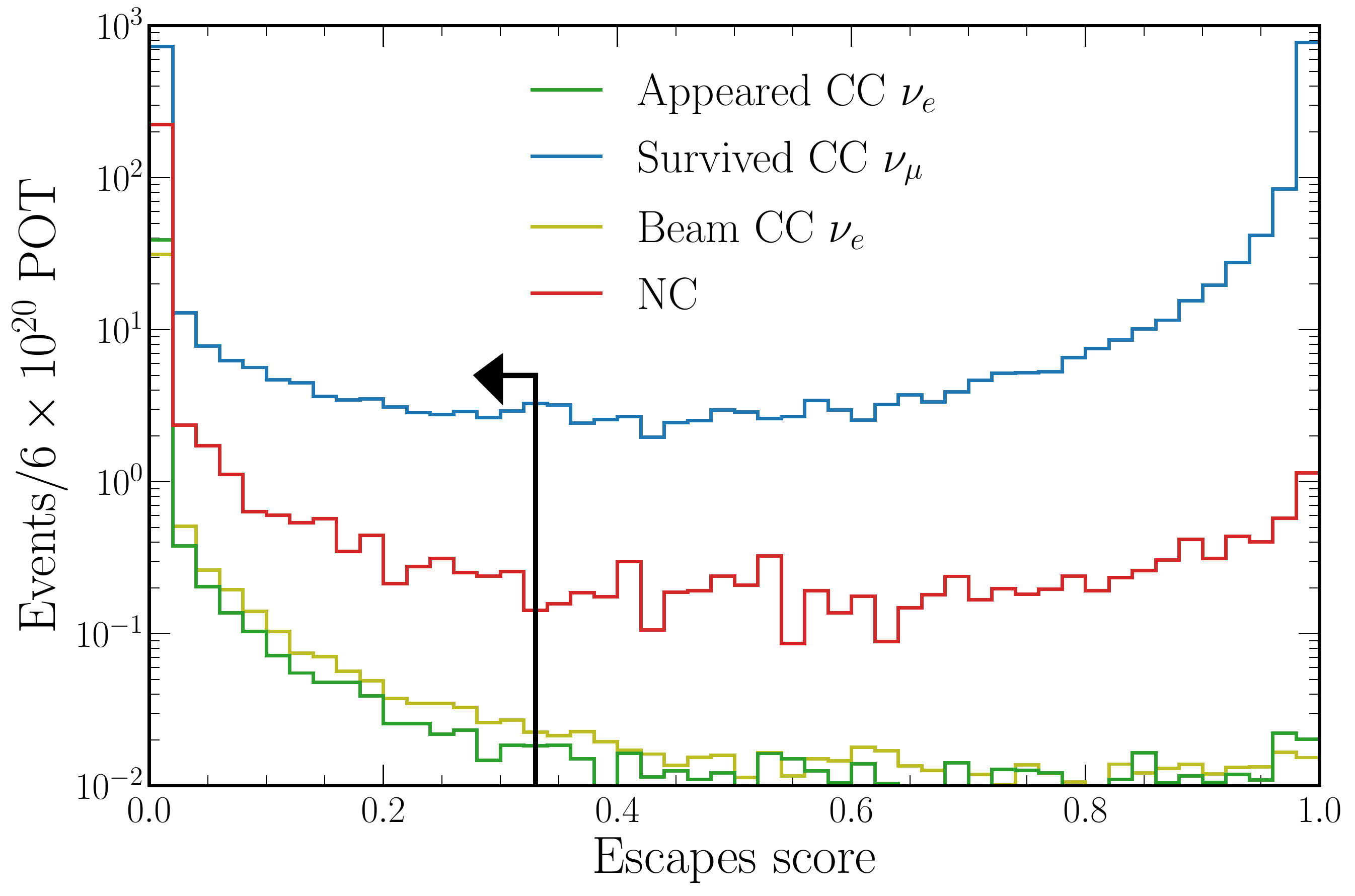}
    \caption{Distribution of \emph{escapes score} output values from the trained cosmic rejection network for the different event categories. A score close to one signifies an escaped primary charged lepton like event, while a score close to zero corresponds to a contained primary charged lepton like event. The containment cut value is shown at 0.33 with the arrow indicating the events that are selected.}
    \label{fig:final_escapes_outputs}
\end{figure}

Therefore, the second output from the cosmic rejection network, \emph{escapes score} is also used to select events. Although this output only considers the primary charged lepton, instead of all event particles, it still acts as a reasonable proxy for event containment. The distribution of \emph{escapes score} output values for each event category is shown in
Figure~\ref{fig:final_escapes_outputs}.

An \emph{escapes score} value below $0.33$ is chosen to select events for which the primary
charged lepton is deemed to be fully contained within the detector. This cut value is chosen to maximise the fraction of CC $\nu_{\mu}$ events which are correctly classified as having their primary charged lepton contained or not within the detector, leading to $96.8\pm0.1\%$ of CC $\nu_{\mu}$ events being classified correctly. As expected, the vast majority ($97.2\pm0.1\%$) of the short track CC $\nu_{e}$ and NC events are selected by this cut.

The total number of expected events per year that pass each successive cut (including preselection) for each event category is shown in Table~\ref{tab:selection}. Both CC $\nu_{e}$ categories are selected with an efficiency greater than 92\%, relative to the total number of expected events, while CC $\nu_{\mu}$ events have a $38.9\pm0.1\%$ selection efficiency, mainly due to the \emph{escapes score} cut (as expected). NC events are found to be primarily rejected by the preselection, while cosmic events are heavily rejected by both the preselection and \emph{cosmic score} cuts.

An upper limit of $<6\pm0.6$ cosmic muon events per year passing all the cuts is determined, showing that the cosmic muon rejection works well. Crucially, of all the true cosmic muon events with a \emph{cosmic score} less than $0.9$, none would be classified as signal CC $\nu_{e}$ events by the beam classification detailed in Section~\ref{sec:results_nuel}. Therefore, the expected cosmic muon contamination of both beam selections (CC $\nu_{e}$ and CC $\nu_{\mu}$) is expected to be negligible relative to the selected number of signal events for each and ignored for the rest of this evaluation.

\begin{table}
    \resizebox{\textwidth}{!}{%
    \begin{tabular}{lrrrrr}
        Selection      & App CC $\nu_{e}$ & CC $\nu_{\mu}$ & Beam CC $\nu_{e}$ & NC & Cosmic \\
        \midrule
        Total events   & $44.17\pm0.02$ & $2045.9\pm33.3$ & $35.06\pm0.01$ & $354.7\pm5.8$ & $2100000\pm200000$ \\
        + Preselection & $41.21\pm0.02$ & $1889.5\pm30.9$ & $33.52\pm0.01$ & $243.2\pm4.0$ & $430000\pm40000$ \\
        + Cosmic cut   & $41.10\pm0.02$ & $1874.4\pm30.7$ & $33.35\pm0.01$ & $241.6\pm4.0$ & $<6\pm0.6$ \\
        + Escapes cut  & $40.68\pm0.02$ & $795.7\pm12.3$  & $32.86\pm0.01$ & $233.0\pm3.9$ & $<6\pm0.6$ \\
        \midrule
        Cuts Eff     & $92.1\pm0.1\%$ & $38.9\pm0.1\%$  & $93.7\pm0.1\%$ & $65.7\pm0.3\%$ & $<2.9\pm0.3\times10^{-6}$ \\
    \end{tabular}
    }
    \caption[Number of events passing successive selection cuts for each event category]
    {The total number of expected (weighted) events and the number that pass successive selection cuts for the different event categories. The preselection, \emph{cosmic score} cut, and \emph{escapes score} cut numbers are shown. The selection efficiency relative to the total number of events after all the cuts have been applied is also shown for each event category. As none out of the $350000$ cosmic events in the evaluation sample are selected after the \emph{cosmic score} cut, an upper limit on the values is instead given.}
    \label{tab:selection}
\end{table}

\subsection{CC $\nu_{e}$ selection}
\label{sec:results_nuel}

\begin{figure} 
    \centering
    \includegraphics[width=0.7\textwidth]{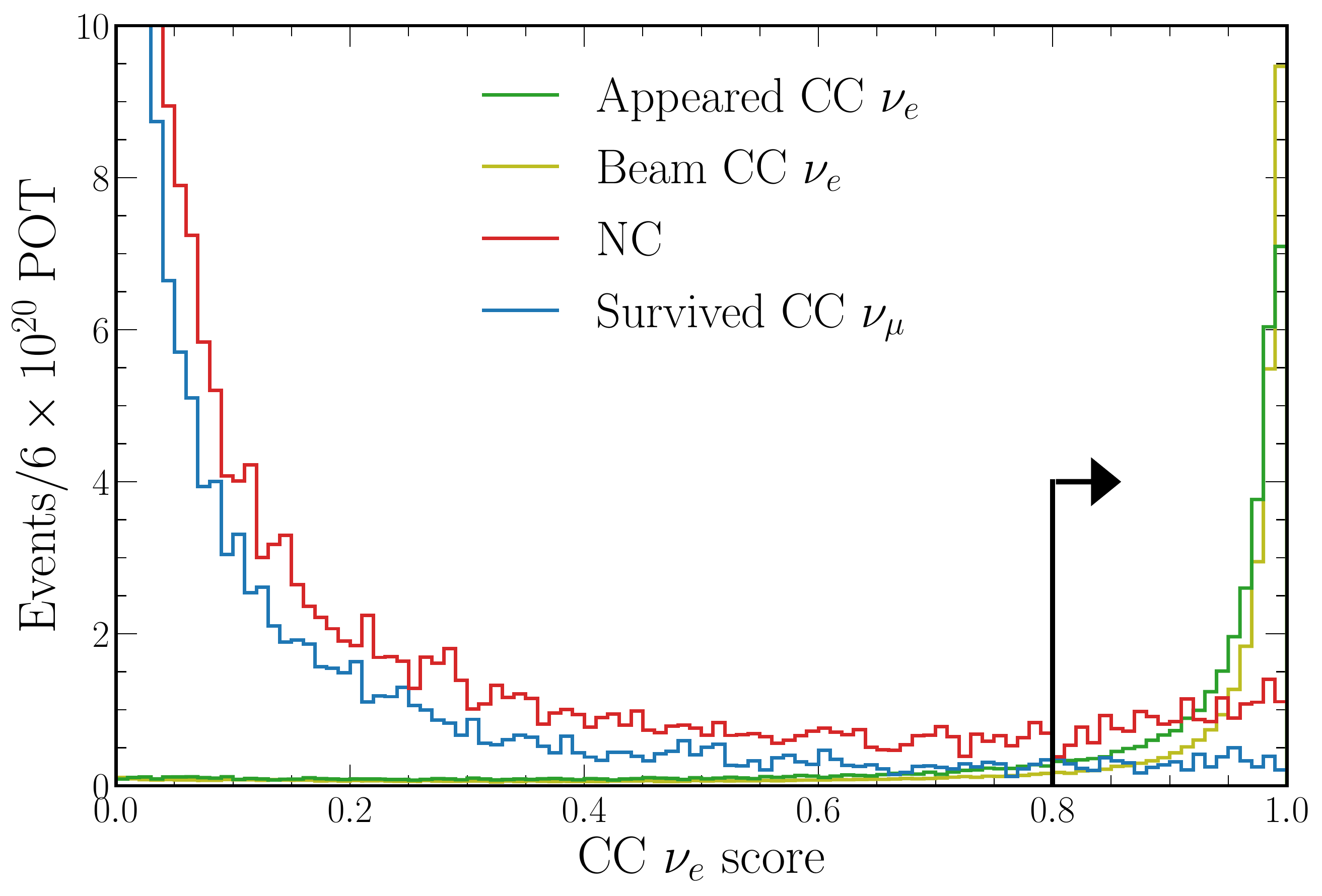}
    \caption{Distribution of \emph{combined category} CC $\nu_{e}$ scores from the trained beam classification network for the different event categories. A score close to one signifies a CC $\nu_{e}$ like event. The FOM-$\nu_e$ optimised cut value is shown at $0.8$ with the arrow indicating the events that are selected. The y-axis has been truncated so that the CC $\nu_{\mu}$ and NC components are not fully visible to better show the distribution of signal CC $\nu_{e}$ events.}
    \label{fig:final_beam_nuel_outputs}
\end{figure}

The distribution of CC $\nu_{e}$ scores from the trained beam classification network form for the different event categories are shown in Figure~\ref{fig:final_beam_nuel_outputs}. A strong separation between appeared CC $\nu_{e}$ signal and both CC $\nu_{\mu}$ and NC background events is achieved. As no attempt is made to separate the appeared CC $\nu_{e}$ signal component from the intrinsic beam CC $\nu_{e}$ background, both are clustered with scores close to one as expected.

To assess the classification performance more rigorously, a selection score for each of the output categories, found by maximising a figure-of-merit (FOM), is calculated. All events with a score above this optimised value are then deemed signal. To minimise the expected measurement statistical error, the value of $\text{efficiency}\times\text{purity}$ (proportional to the square of $s/\sqrt{s+b}$) is optimised as the FOM~\cite{list2002}.

\begin{figure} 
    \centering
    \includegraphics[width=0.7\textwidth]{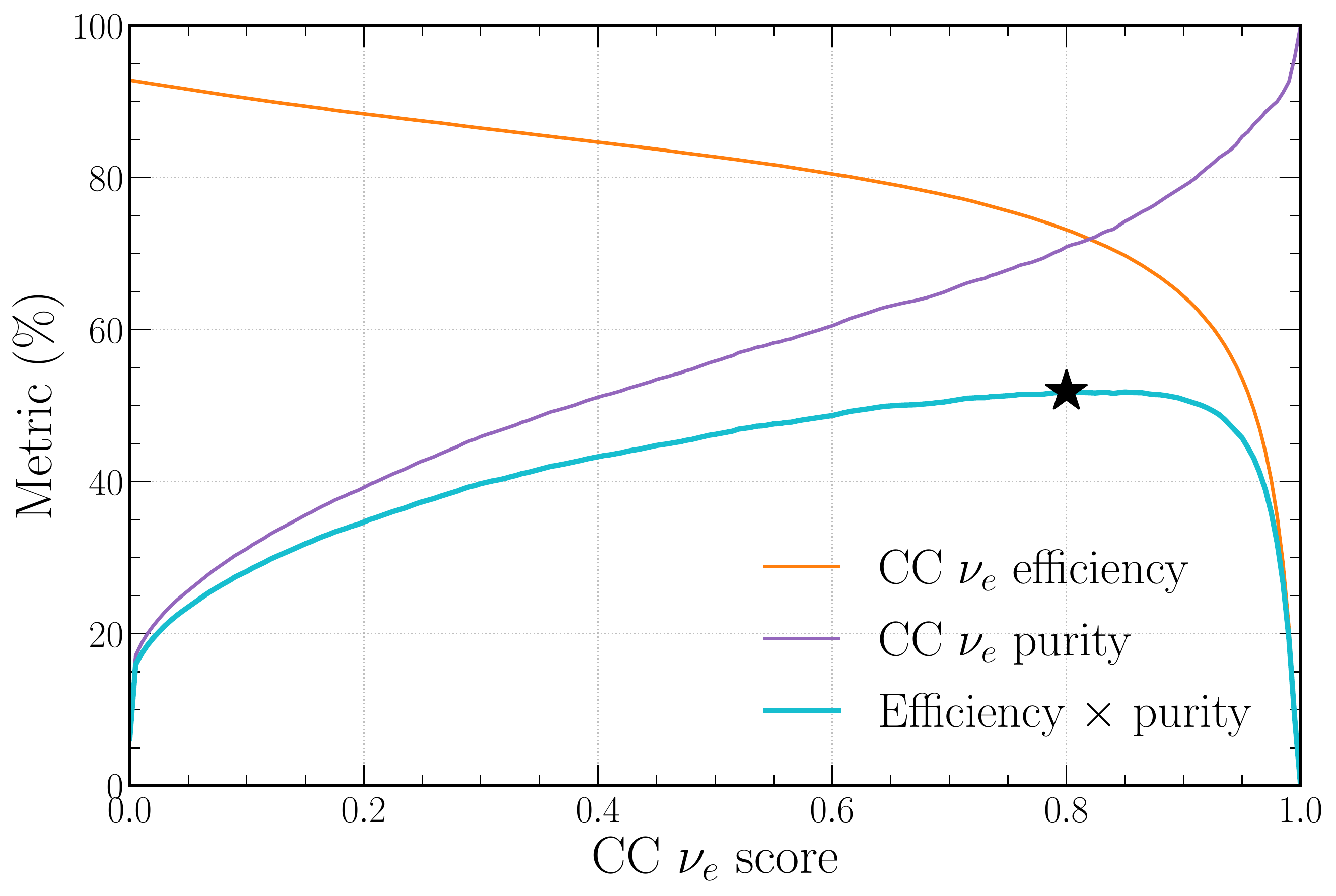}
    \caption{CC $\nu_{e}$ efficiency, purity, and $\text{efficiency}\times\text{purity}$ (FOM-$\nu_e$) curves for different values of CC $\nu_{e}$ score selection. The Maximum FOM-$\nu_{e}$ value of $0.519\pm0.004$ is indicated at a CC $\nu_{e}$ score of $0.8$ by the star.}
    \label{fig:final_nuel_eff_curves}
\end{figure}

The efficiency, purity, and their product (the FOM-$\nu_e$) for CC $\nu_{e}$ events (both appeared and beam) as a function of selecting events above a certain CC $\nu_{e}$ score are shown in Figure~\ref{fig:final_nuel_eff_curves}. The FOM-$\nu_e$ is optimised by selecting events with a CC $\nu_{e}$ score above $0.8$, achieving a value of $0.519\pm0.004$. Note that the FOM-$\nu_e$ is optimised considering both the appeared and beam CC $\nu_{e}$ components as signal due to their indistinguishable nature.

The total number of events are shown in Table~\ref{tab:nuel_selection} with event category alongside the corresponding selection efficiencies and appeared CC $\nu_{e}$ signal and combined CC $\nu_{e}$ purities. The purities are defined as the fraction of events within the selection that are true signal events. The final FOM-$\nu_e$ selected appeared CC $\nu_{e}$ signal purity of $38.3\pm0.3\%$ may appear low, but this is mainly due to the indistinguishable intrinsic beam CC $\nu_{e}$ contamination, note that when both CC $\nu_{e}$ components are considered signal, the selection purity is $70.9\pm0.6\%$.

\begin{table}
    \resizebox{\textwidth}{!}{%
    \begin{tabular}{lrrrrrr}
        Selection & CC $\nu_{e}$ sig & CC $\nu_{\mu}$ bkg & CC $\nu_{e}$ bkg & NC bkg & Purity sig
        & Purity CC $\nu_{e}$ \\
        \midrule
        Total events    & $44.17\pm0.02$ & $2045.9\pm33.3$ & $35.06\pm0.01$ & $354.7\pm5.8$ & $1.78\pm0.02\%$ & $3.19\pm0.03\%$ \\
        + Cuts          & $40.68\pm0.02$ & $795.7\pm12.3$  & $32.86\pm0.01$ & $233.0\pm3.9$ & $3.69\pm0.02\%$ & $6.67\pm0.03\%$ \\
        + FOM-$\nu_e$   & $31.27\pm0.02$ & $6.0\pm0.1$     & $26.69\pm0.01$ & $17.8\pm0.3$  & $38.3\pm0.3\%$  & $70.9\pm0.6\%$ \\
        \midrule
        Cuts Eff        & $92.1\pm0.1\%$ & $38.9\pm0.1\%$  & $93.7\pm0.1\%$ & $65.7\pm0.3\%$ & - & - \\
        FOM-$\nu_e$ Eff & $70.8\pm0.2\%$ & $0.29\pm0.02\%$ & $76.1\pm0.1\%$ & $5.0\pm0.2\%$  & - & - \\
    \end{tabular}
    }
    \caption[Table showing CC $\nu_{e}$ selected event numbers, efficiencies and purities]
    {Table showing CC $\nu_{e}$ selected event numbers and corresponding efficiencies for the various event categories as well as associated purities. Shown are the total event
    numbers, those after the preselection, \emph{cosmic score} cut, and \emph{escapes score} cut (Cuts), in addition to the numbers after the FOM-$\nu_e$ optimised selection with the efficiencies relative to the total number of events shown for both. Both the appeared signal CC $\nu_{e}$ purity and the joint appeared and beam CC $\nu_{e}$ purity are shown for each selection.}
    \label{tab:nuel_selection}
\end{table}

\begin{figure} 
    \centering
    \includegraphics[width=0.7\textwidth]{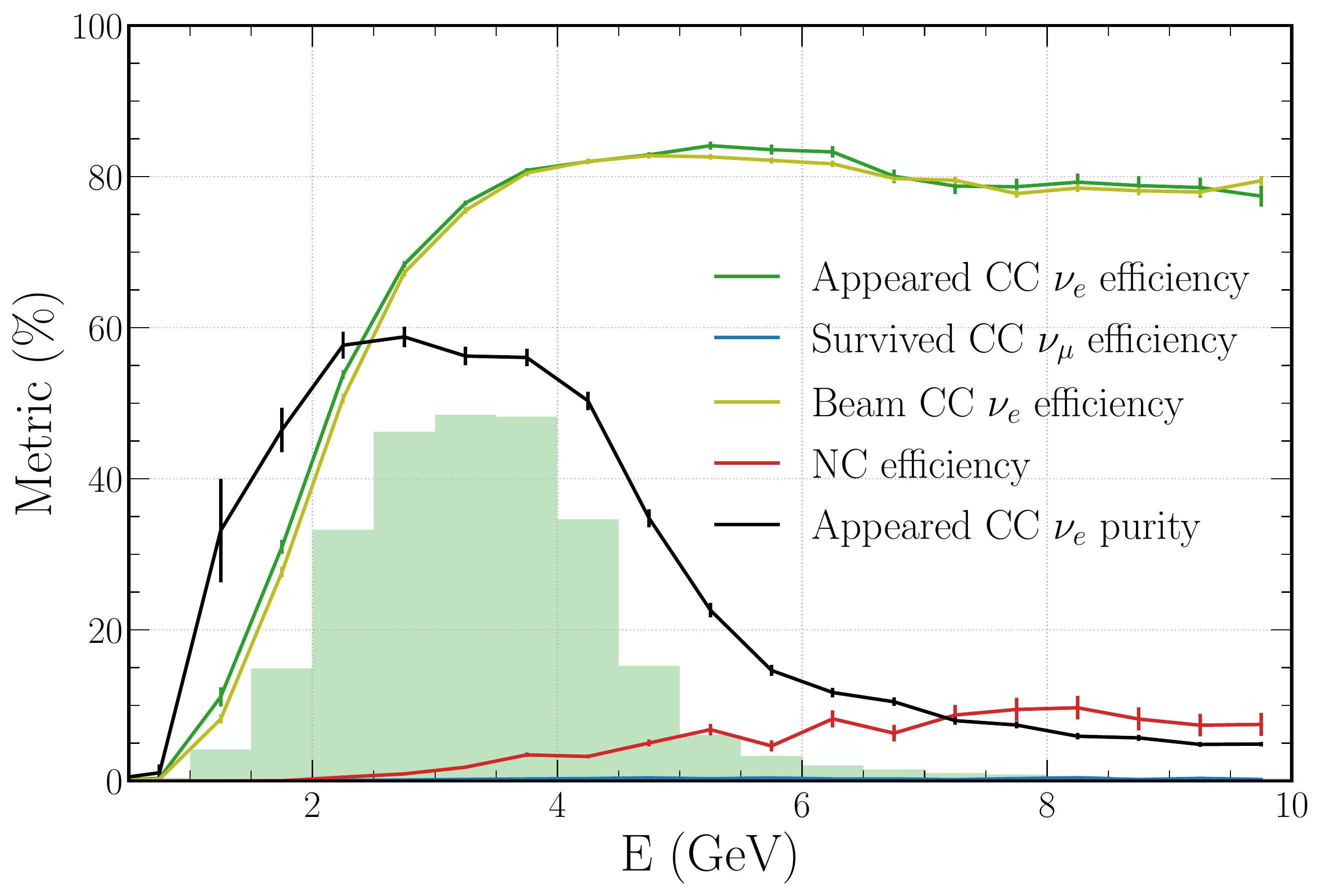}
    \caption{FOM-$\nu_e$ selection efficiencies relative to the total number of events for the different event categories as well as appeared CC $\nu_{e}$ purity as a function of energy. All CC categories are shown in terms of the true neutrino energy, while NC events are shown in terms of the true hadronic component energy. The survived CC $\nu_{\mu}$ efficiency is so low it is barely visible near zero. For reference, the true appeared CC $\nu_{e}$ neutrino energy distribution is shown in green.}
    \label{fig:final_nuel_hists}
\end{figure}

The FOM-$\nu_e$ optimised CC $\nu_{e}$ selection efficiency, relative to the total number of events as a function of energy for the different event categories, is shown in Figure~\ref{fig:final_nuel_hists} alongside the signal purity. From low neutrino energies, both CC $\nu_{e}$ category selection efficiencies rise to a plateau of approximately 80\% beginning at \SI{4}{\GeV}. This is expected as low energy CC $\nu_{e}$ events have less well-defined electron Cherenkov rings, leading to their rejection. Problematically, this turn-on behaviour cuts into the true appeared CC $\nu_{e}$ distribution, especially around the \SI{1.5}{\GeV} oscillation maximum. Future work should explore whether a CC $\nu_{e}$ selection cut that varies with energy can lead to a greater proportion of these low energy events being selected.

Due to the abundance of selected intrinsic beam CC $\nu_{e}$ events at higher energies, the appeared CC $\nu_{e}$ purity is observed to peak at approximately \SI{2.5}{\GeV} (reasonably close to the oscillation maximum) before declining. Importantly, within the key signal region from $2$ to \SI{4}{\GeV}, the appeared CC $\nu_{e}$ purity is $>55\%$, larger than the $38.3\pm0.3\%$ across the full evaluation sample. The NC efficiency is seen to slowly increase, approaching 15\% for hadronic component energies above \SI{5}{\GeV}; this is likely due to misidentification of high energy pions or protons as electrons. Crucially, however, within the key signal region, NC selection efficiency remains low.

The best way to understand the relative performance of the CNN CC $\nu_{e}$ classification is by comparison with the standard event selection, outlined in detail in references~\cite{jtingey2021} and~\cite{blake2016}. A maximum $\text{efficiency}\times\text{purity}$ of $0.132\pm0.005$ is achieved; only $\sim25\%$ the value reached by the CNN approach. Both the combined appeared and beam CC $\nu_{e}$ efficiency of $34.7\pm0.8\%$ compared to $73.4\pm0.2\%$ and purity of $39.3\pm1.2\%$ compared to $70.9\pm0.6\%$ are considerably lower than that provided by the new CNN classification.

Furthermore, the new appeared CC $\nu_{e}$ signal efficiency of $70.8\pm0.2\%$ compares well to the 62\% and 64\% achieved by the \nova and T2K CC $\nu_{e}$ selections, respectively. However, purity is significantly lower at $38.3\pm0.3\%$ compared to the 78\% and 80\% reached by \nova and T2K~\cite{acero2019, abe2015}. A large proportion of this difference can be explained by the lower neutrino energies and greater off-axis angles at which these experiments operate. Not only does this increase the proportion of easy to identify CC-QEL events, but it also reduces the indistinguishable beam CC $\nu_{e}$ contamination.

\subsection{CC $\nu_{\mu}$ selection}
\label{sec:results_numu}
\begin{figure} 
    \centering
    \includegraphics[width=0.7\textwidth]{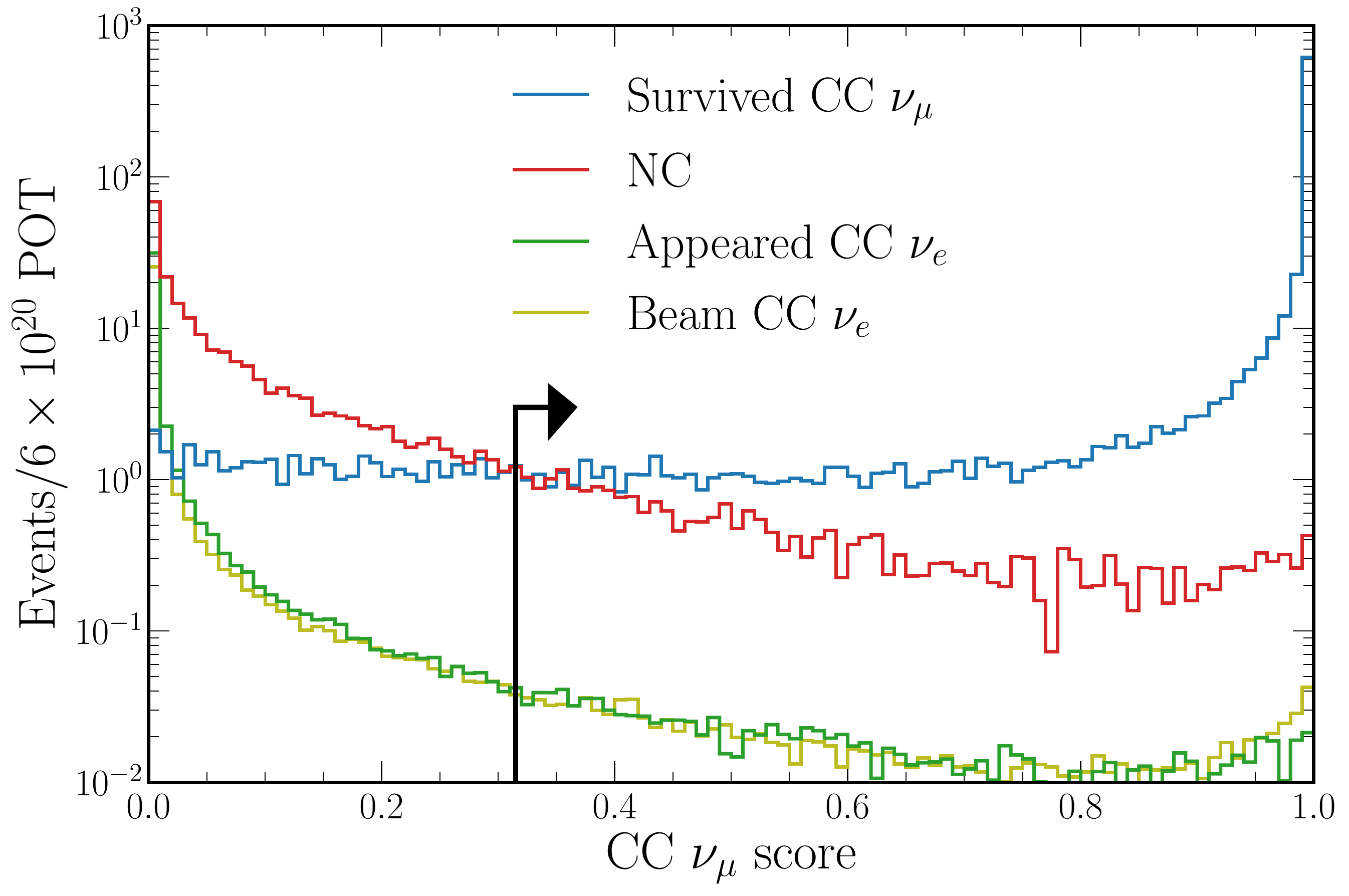}
    \caption{Distribution of \emph{combined category} CC $\nu_{\mu}$ scores from the trained beam classification network for the different event categories. A score close to one signifies a CC $\nu_{\mu}$ like event. The FOM-$\nu_{\mu}$ optimised cut value is shown at $0.315$ with the arrow indicating the events that are selected.}
    \label{fig:final_beam_numu_outputs}
\end{figure}

The distribution of CC $\nu_{\mu}$ scores from the trained beam classification network form for the different event categories are shown in Figure~\ref{fig:final_beam_numu_outputs}. Excellent separation between appeared CC $\nu_{\mu}$ signal and both CC $\nu_{e}$ components and NC background is achieved. For high CC $\nu_{\mu}$ scores (close to one) the difference between signal and background rates is approximately three orders of magnitude.

\begin{figure} 
    \centering
    \includegraphics[width=0.7\textwidth]{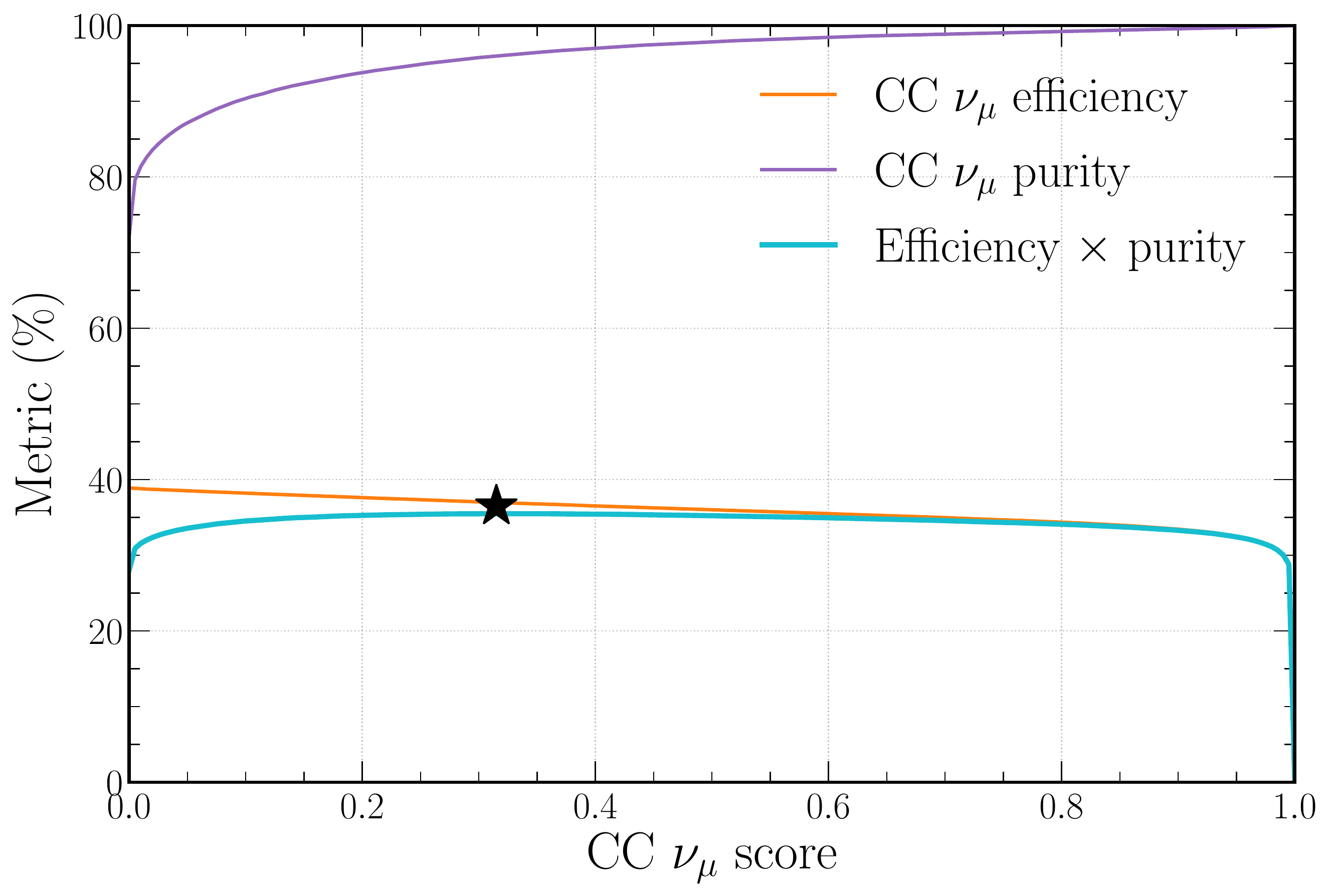}
    \caption{CC $\nu_{\mu}$ efficiency, purity, and $\text{efficiency}\times\text{purity}$ (FOM-$\nu_{\mu}$) curves for different values of CC $\nu_{\mu}$ score selection. The maximum FOM-$\nu_{\mu}$ value of $0.365\pm0.002$ is indicated at a CC $\nu_{\mu}$ score of $0.315$ by the star.}
    \label{fig:final_numu_eff_curves}
\end{figure}

The efficiency, purity, and their product (the FOM-$\nu_{\mu}$) for CC $\nu_{\mu}$ events as a function of selecting events above a certain CC $\nu_{\mu}$ score are shown in Figure~\ref{fig:final_numu_eff_curves}. FOM-$\nu_{\mu}$ is optimised by selecting events with a CC $\nu_{\mu}$ score above $0.315$, achieving a value of $0.365\pm0.002$. The total number of events, those selected by the previously mentioned cuts, and those furthermore selected by the FOM-$\nu_{\mu}$ optimised selection are shown in Table~\ref{tab:numu_selection} for each event category alongside the corresponding selection efficiencies and CC $\nu_{\mu}$ signal purity.

The signal efficiency of $37.0\pm0.1\%$ compares well to the 31\% and 36\% achieved by the \nova and T2K CC $\nu_{\mu}$ selections, respectively~\cite{acero2019, abe2015}. This is also the case for the signal purity of $96.0\pm0.1\%$ compared to the 98.6\% and 94\% purities of the \nova and T2K selections. Although the final signal efficiency is low, this is desirable to ensure events are fully contained for energy estimation. When considering just those CC $\nu_{\mu}$ events for which the primary charged muon is contained within the detector volume at the truth level, an $87.5\pm0.1\%$ selection efficiency is achieved.

\begin{table}
    \resizebox{\textwidth}{!}{%
    \begin{tabular}{lrrrrr}
        Selection & CC $\nu_{\mu}$ sig & App CC $\nu_{e}$ bkg & Beam CC $\nu_{e}$ bkg & NC bkg &
        Purity sig \\
        \midrule
        Total events & $2045.9\pm33.3$ & $44.17\pm0.02$ & $35.06\pm0.01$ & $354.7\pm5.8$ & $82.5\pm0.2\%$ \\
        + Cuts  & $795.7\pm12.3$  & $40.68\pm0.02$ & $32.86\pm0.01$ & $233.0\pm3.9$ & $72.2\pm0.2\%$ \\
        + FOM-$\nu_{\mu}$   & $756.4\pm11.6$  & $1.293\pm0.001$ & $1.315\pm0.001$ & $29.0\pm0.5$ & $96.0\pm0.1\%$ \\
        \midrule
        Cuts Eff        & $38.9\pm0.1\%$ & $92.1\pm0.1\%$ & $93.7\pm0.1\%$ & $65.7\pm0.3\%$ & - \\
        FOM-$\nu_{\mu}$ Eff & $37.0\pm0.1\%$ & $2.9\pm0.1\%$  & $3.8\pm0.1\%$   & $8.2\pm0.2\%$ & - \\
    \end{tabular}
    }
    \caption[Table showing CC $\nu_{\mu}$ selected event numbers, efficiencies and signal purity]
    {Table showing CC $\nu_{\mu}$ selected event numbers and corresponding efficiencies for the various event categories as well as the associated signal purity. Shown are the total event numbers, those after the preselection, \emph{cosmic score} cut, and \emph{escapes score} cut (Cuts), in addition to the numbers after the FOM-$\nu_{\mu}$ optimised selection with the efficiencies relative to the total number of events shown for both. The CC $\nu_{\mu}$ signal purity is also shown for each selection.}
    \label{tab:numu_selection}
\end{table}

\begin{figure} 
    \centering
    \includegraphics[width=0.7\textwidth]{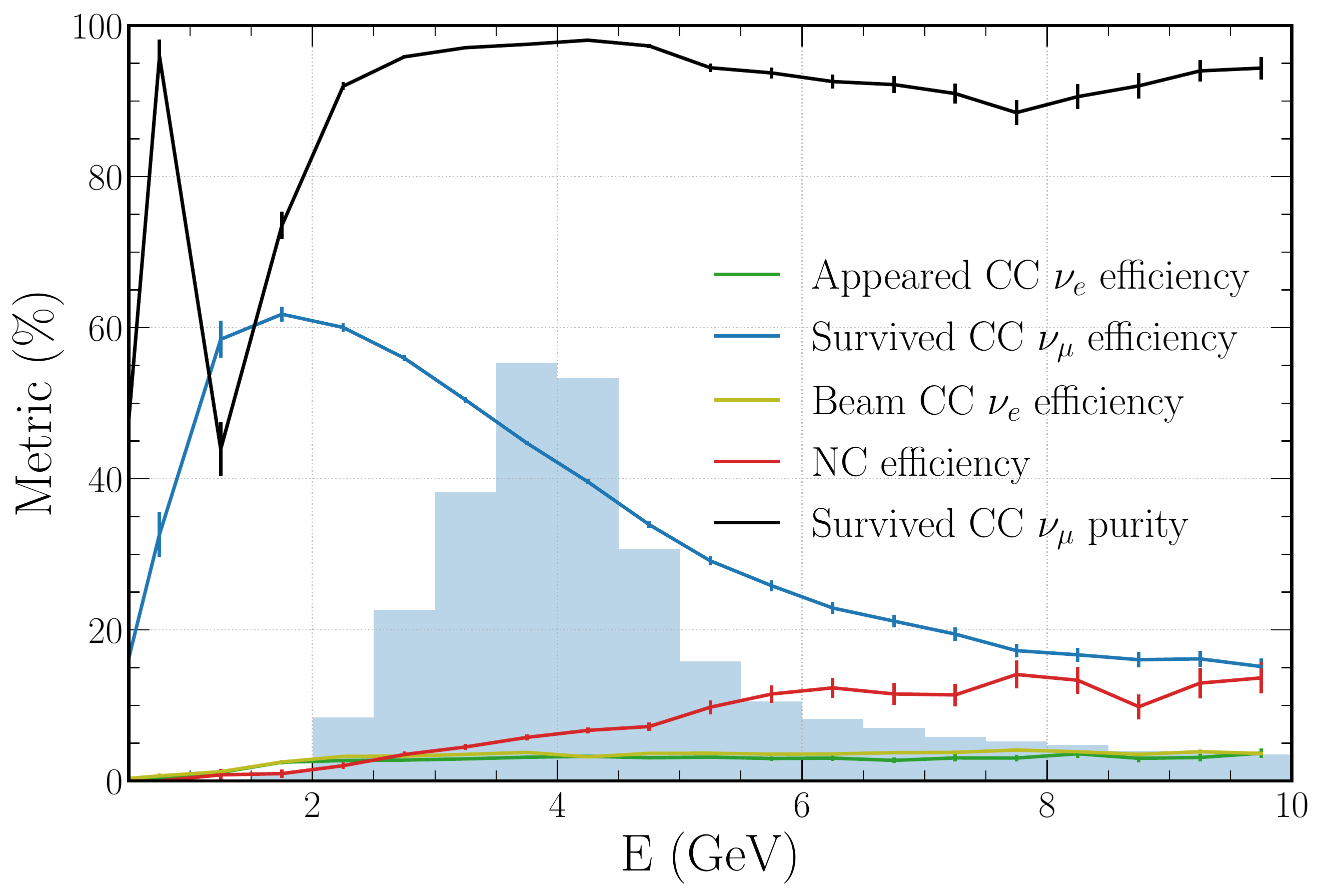}
    \caption{FOM-$\nu_{\mu}$ selection efficiencies relative to the total number of events for the different event categories as well as the survived CC $\nu_{\mu}$ purity as a function of energy. All CC categories are shown in terms of the true neutrino energy, while NC events are shown in terms of the true hadronic component energy. For reference, the true survived CC $\nu_{\mu}$ neutrino energy distribution is shown in blue.}
    \label{fig:final_numu_hists}
\end{figure}

The FOM-$\nu_{\mu}$ optimised CC $\nu_{\mu}$ selection efficiency, relative to the total number of events as a function of energy for the different event categories, is shown in
Figure~\ref{fig:final_numu_hists}. Survived CC $\nu_{\mu}$ selection efficiency peaks at just below \SI{2}{\GeV} before slowly declining, this is explained by higher energy events being less likely to have their primary charged muon fully contained within the detector. Of interest is the expected dip in the otherwise very high ($>90\%$) CC $\nu_{\mu}$ purity at approximately \SI{1.5}{\GeV}, corresponding to approximately the muon neutrino oscillation maximum. As in the CC $\nu_{e}$ selection case, the NC efficiency is seen to rise with energy, again likely due to misidentification of energetic protons and pions.

\subsection{Interaction type classification} 
\label{sec:results_interaction}

\begin{figure} 
    \centering
    \includegraphics[width=0.8\textwidth]{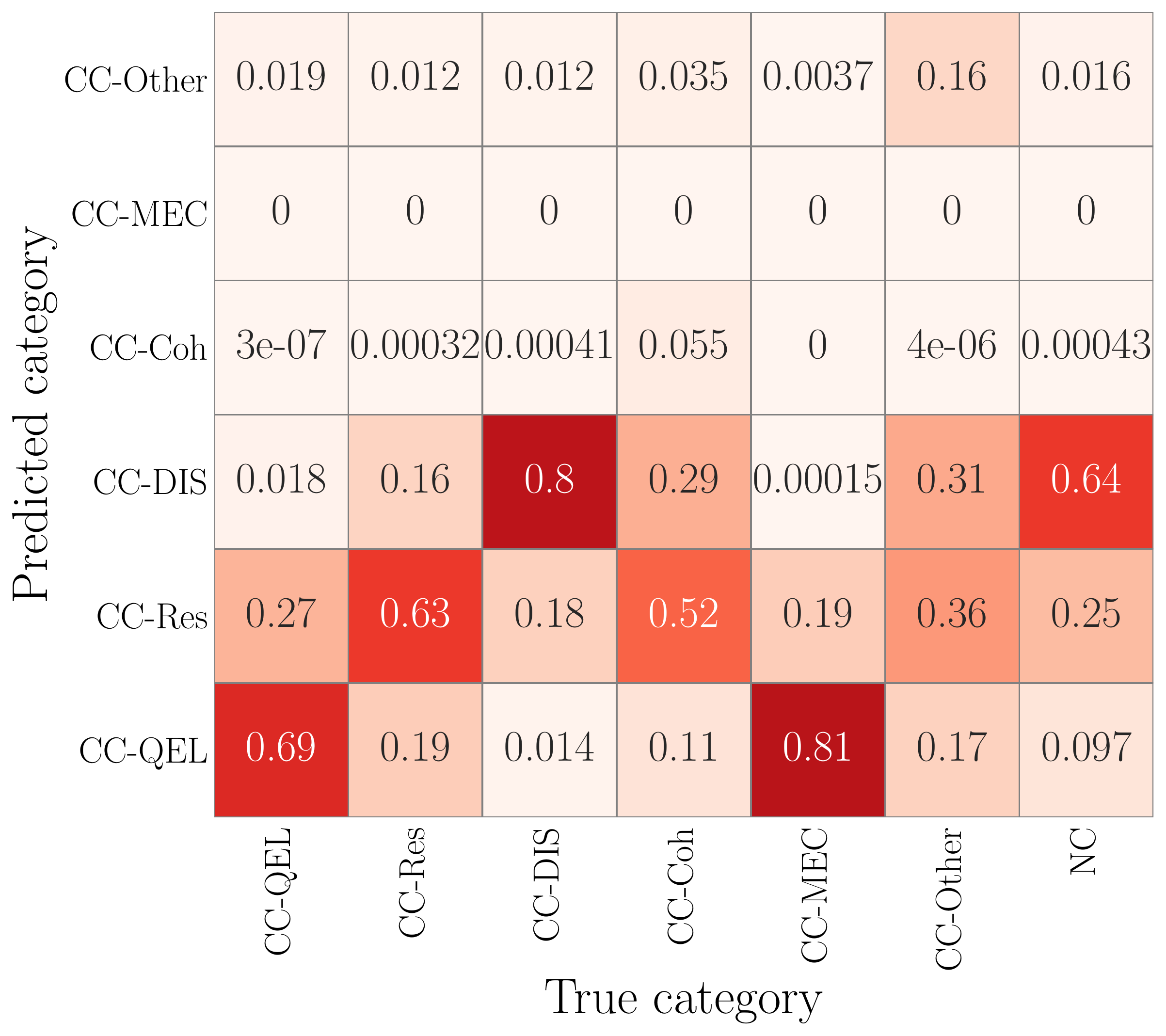}
    \caption[Classification matrix for the CC category output of the beam classification network]
    {Classification matrix for the \emph{CC category} output of the trained beam classification
    network. Shown are events that have either been selected by the CC $\nu_{e}$ or CC $\nu_{\mu}$ selection. Events are simply classified using the categorical score for which they have the highest value. The numbers shown are the fraction of true category events classified into each of the six possible categories.}
    \label{fig:final_cc_cat_confusion}
\end{figure}

Using the \emph{CC category} output of the trained beam classification network form, the CC interaction type for both CC $\nu_{e}$ and CC $\nu_{\mu}$ selected events can be determined. As in the \emph{combined category} output case, the highest-scoring neuron can be used for classification, resulting in the matrix shown in Figure~\ref{fig:final_cc_cat_confusion}. Note that only events which are selected by either the CC $\nu_{e}$ or CC $\nu_{\mu}$ selection are shown.

Reasonable classification accuracy greater than 60\% is achieved across the three dominant interaction types, CC-QEL, CC-Res, and CC-DIS. The less common CC-Coh and CC-MEC types are found to be commonly misidentified as CC-Res and CC-QEL respectively, likely due to the imbalanced training dataset and their corresponding topological similarities. Background NC events which pass either CC $\nu_{e}$ or CC $\nu_{\mu}$ selection (commonly high in energy) are found to be typically classified as CC-DIS; this is expected as they commonly contain multiple energetic particles in the final state.

\subsection{Energy estimation}
\label{sec:results_energy}
By using the CC interaction type classification just described, the differences between CC interaction types can be exploited to improve neutrino energy, charged lepton energy, and interaction vertex position and time estimation. For events classified as either CC $\nu_{e}$ or CC $\nu_{\mu}$ with an associated \emph{CC category} interaction type, the corresponding bespoke energy estimation network form outlined in Section~\ref{sec:cnn_energy_training} is used for estimation.

Only three networks for each neutrino type are trained, one for each of the dominant interaction types CC-QEL (and CC-MEC), CC-Res, and CC-DIS. For events not classified by the \emph{CC category} output as one of these categories, such as CC-Coh or CC-Other, the CC-Res network is used as it is the most topologically similar interaction type.

\begin{figure} 
    \includegraphics[width=\textwidth]{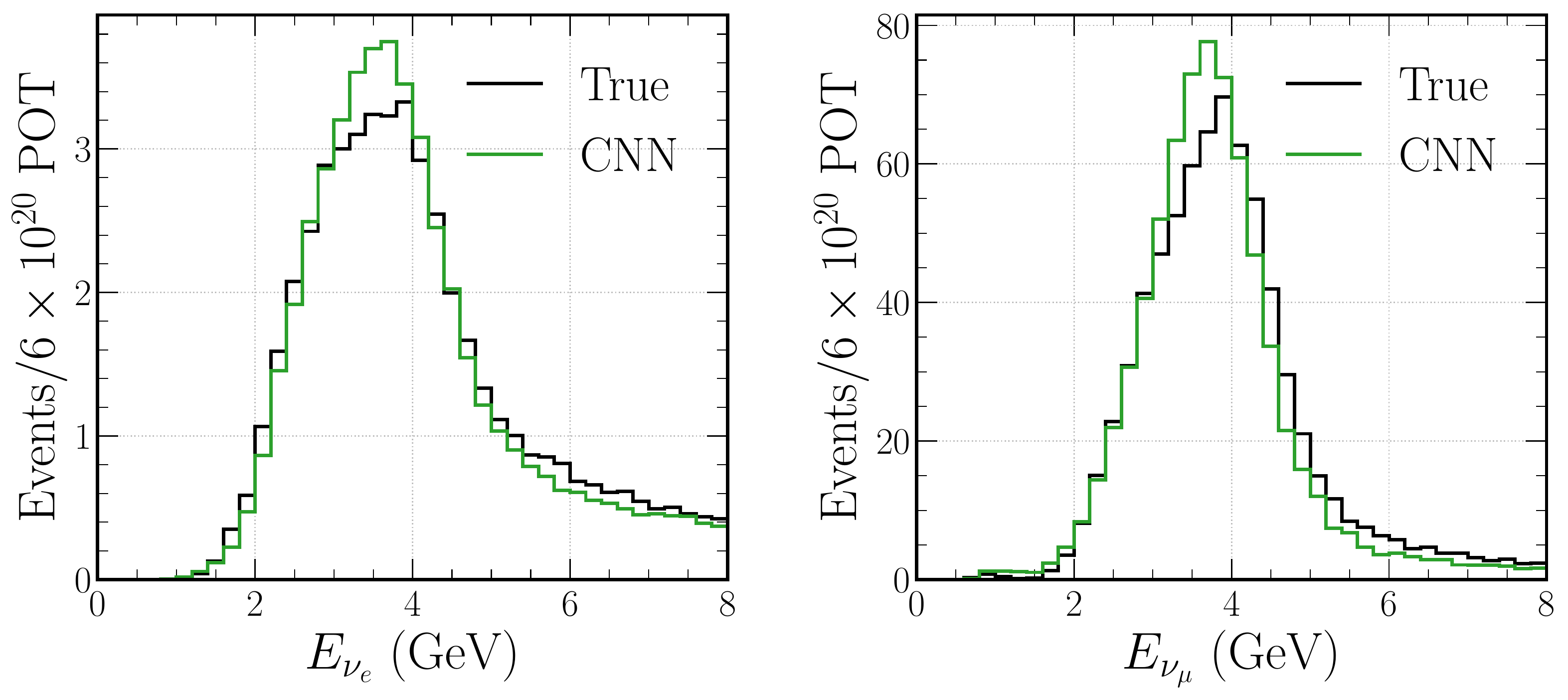}
    \caption{Distributions of true and CNN estimated neutrino energy for CC $\nu_{e}$ (left) and CC $\nu_{\mu}$ (right) beam events. Only true CC $\nu_{e}$ and CC $\nu_{\mu}$ events that are also selected by the CC $\nu_{e}$ and CC $\nu_{\mu}$ selections respectively are shown.}
    \label{fig:final_energy_dists}
\end{figure}

The distributions of CNN estimated \emph{neutrino energy} output and true $\nu_{e}$ and $\nu_{\mu}$ neutrino energies for true CC $\nu_{e}$ and CC $\nu_{\mu}$ events respectively that are also selected by their corresponding CC selection are shown in Figure~\ref{fig:final_energy_dists}. The CNN estimated distributions match the truth well across the full range of neutrino energies expected within \chipsfive, except in the peak regions where the truth distribution shape is not fully captured.

\begin{figure} 
    \includegraphics[width=\textwidth]{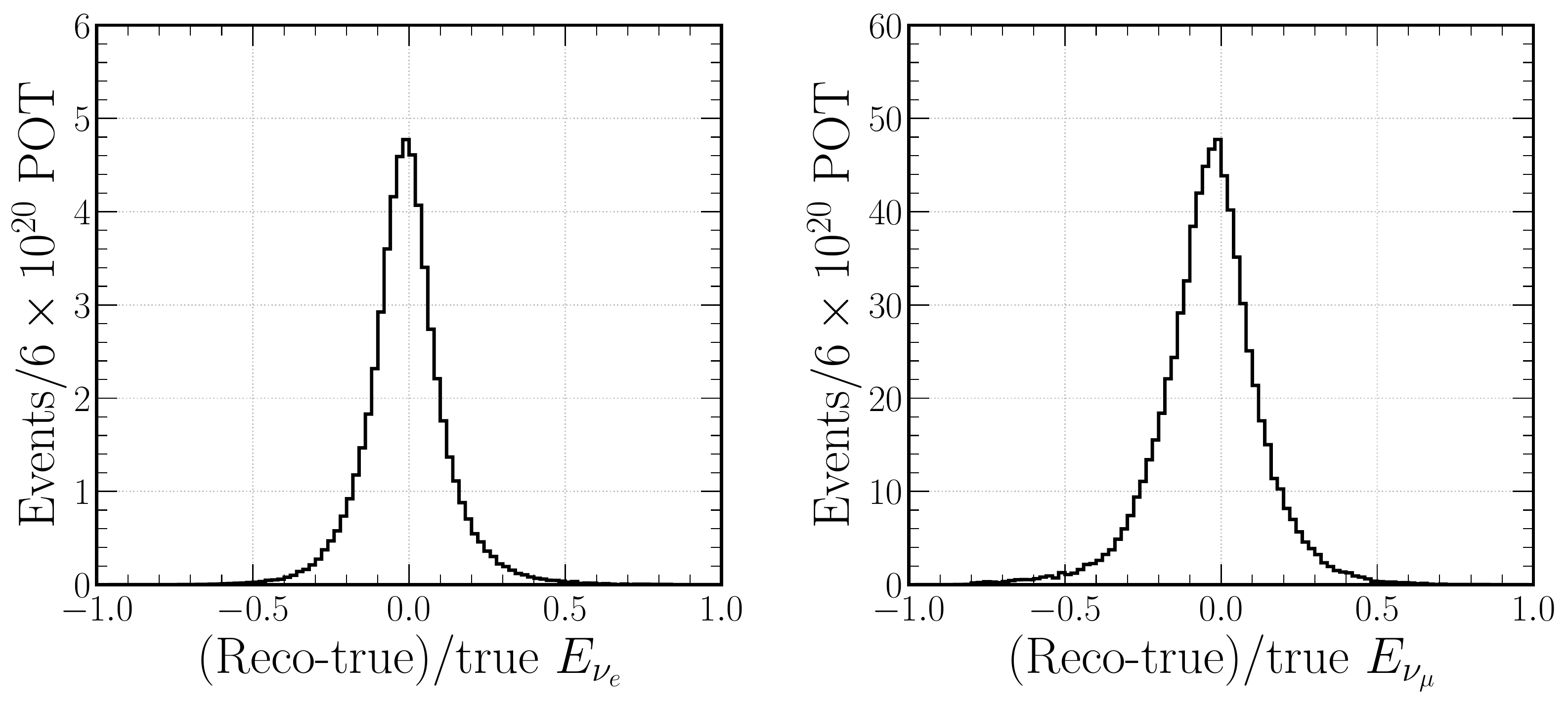}
    \caption{Distributions of (reco-true)/true neutrino energies for both selected CC $\nu_{e}$ (left) and CC $\nu_{\mu}$ (right) signal beam events.}
    \label{fig:final_energy_frac}
\end{figure}

\begin{figure} 
    \includegraphics[width=\textwidth]{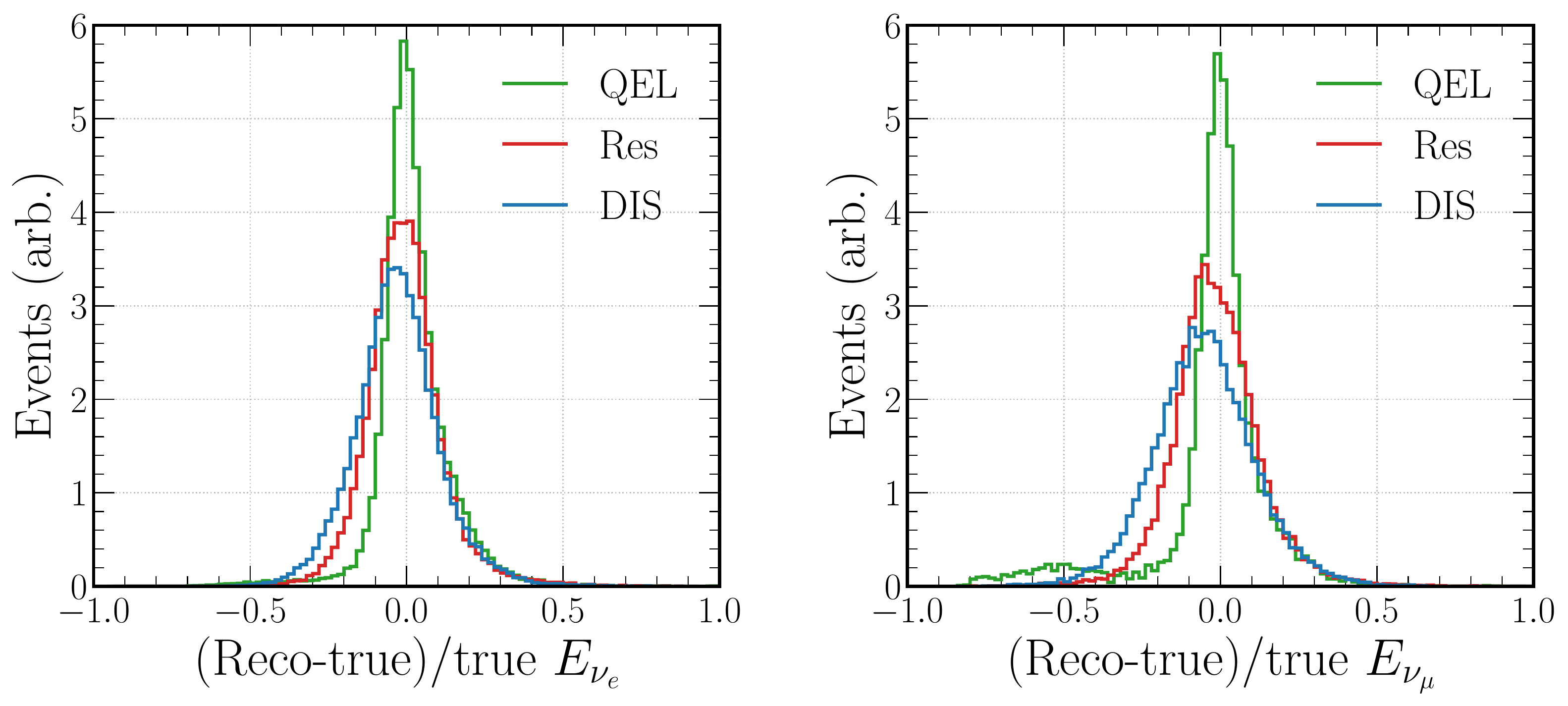}
    \caption{Distributions of (reco-true)/true neutrino energies for both selected CC $\nu_{e}$ (left) and CC $\nu_{\mu}$ (right) signal beam events by interaction type. The relative number of events between interaction types has been scaled for clearer comparison.}
    \label{fig:final_energy_frac_split}
\end{figure}

To fully understand CNN energy estimation performance, histograms of ratios of fractional differences between CNN estimated (reco) and true neutrino energy for both true CC $\nu_{e}$ and CC $\nu_{\mu}$ beam events that are also selected by their corresponding CC selection are shown in Figure~\ref{fig:final_energy_frac}. Similar distributions splitting the signal components by interaction type are shown in Figure~\ref{fig:final_energy_frac_split}. Furthermore, the Full Width Half Maximum (FWHM) neutrino energy resolutions derived from these plots for both CC $\nu_{e}$ and CC $\nu_{\mu}$ events are shown in Table~\ref{tab:energy_resolutions}.

\begin{table}
    \resizebox{\textwidth}{!}{%
    \begin{tabular}{lrrrrr}
        Event type     & All & QEL component & Res component & DIS component \\
        \midrule
        CC $\nu_{e}$   & $24.0\pm0.3\%$ & $16.4\pm0.4\%$ & $24.3\pm0.2\%$ & $31.9\pm0.2\%$ \\
        CC $\nu_{\mu}$ & $29.4\pm0.4\%$ & $14.1\pm0.3\%$ & $27.2\pm0.3\%$ & $34.9\pm0.3\%$ \\
    \end{tabular}
    }
    \caption[Summary of CC $\nu_{e}$ and CC $\nu_{\mu}$ FWHM resolutions]
    {Summary of CC $\nu_{e}$ and CC $\nu_{\mu}$ FWHM neutrino energy resolutions. Shown for each sample are the FWHM values for all selected signal events and the three dominant
    interaction type components, QEL, Res, and DIS. The FWHM values are calculated from the
    distributions shown in Figure~\ref{fig:final_energy_frac} and Figure~\ref{fig:final_energy_frac_split}.}
    \label{tab:energy_resolutions}
\end{table}

The interaction type FWHM values follow the expected pattern, with the simple to reconstruct single charged lepton QEL interactions achieving a smaller value than multi-particle DIS events. Furthermore, when the approximate resolution is derived from the FWHM values\footnote{The approximate resolution, given by the standard deviation $\sigma$ is found by dividing the FWHM by $2\sqrt{2\ln2}\approx2.355$.} the resolutions of $10.2\pm0.2\%$ and $12.5\pm0.2\%$ for CC $\nu_{e}$ and CC $\nu_{\mu}$ respectively are comparable to the resolutions obtained by \nova of 10.7\% and 9.1\%~\cite{acero2019}.

\begin{figure} 
    \includegraphics[width=\textwidth]{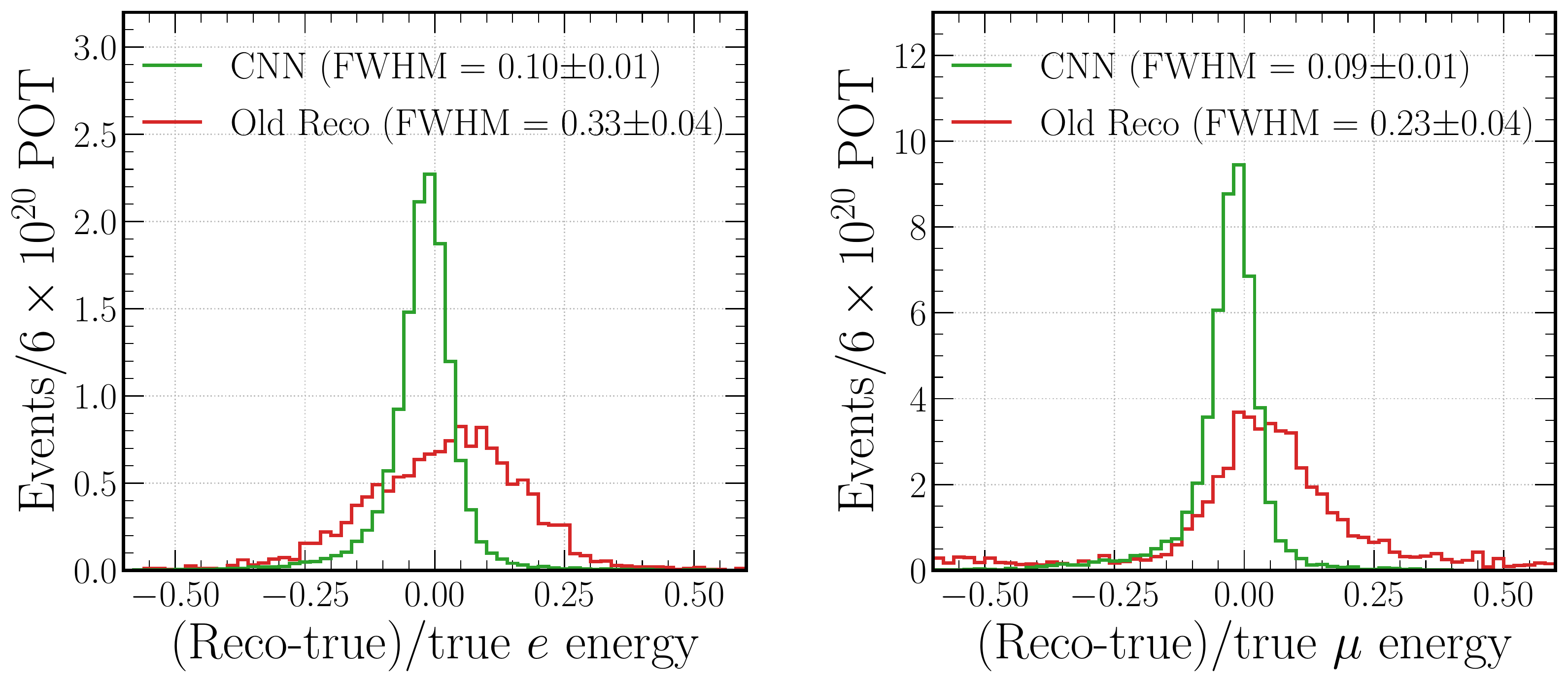}
    \caption[Distributions of (reco-true)/true primary charged lepton energies for the CNN and standard methods] {Distributions of (reco-true)/true primary charged lepton energies for both CC $\nu_{e}$ (left) and CC $\nu_{\mu}$ (right) beam events for both the new CNN approach and standard (old) methods. Only QEL events are shown for clearer comparison with the standard reconstruction methods. The FWHM values of each distribution are indicated in
    the legends.}
    \label{fig:final_frac_e_comparison}
\end{figure}

As in the CC $\nu_{e}$ selection case, the best way to understand the relative performance of the energy estimation is by comparison with the standard \chips reconstruction outlined in detail in references~\cite{jtingey2021} and~\cite{blake2016}. Although the standard reconstruction does not attempt to estimate the neutrino energy, the energy of the primary charged lepton in each CC event is predicted. The value can be compared to the \emph{charged lepton energy} output of the energy estimation CNNs. Histograms of ratios of differences between CNN estimated (reco) and true charged lepton energy to true charged lepton energy for both CC $\nu_{e}$ and CC $\nu_{\mu}$ beam QEL events are shown in Figure~\ref{fig:final_frac_e_comparison}.

A significant improvement is made using the new CNN approach. FWHM lepton energy values 30\% and 39\% the size of the standard reconstruction values for CC $\nu_{e}$ and CC $\nu_{\mu}$ QEL events respectively is achieved, at $10.0\pm0.1\%$ and $9.0\pm0.1\%$, in their fractional percentage form. When the approximate resolution is found from the CC $\nu_{e}$ FWHM value ($4.2\pm0.1\%$), it compares well to the $\sim2.5\%$ CC QEL charged lepton energy resolution reached by the Super-Kamiokande fiTQun algorithm~\cite{jiang2019}. Impressive, given the significant differences in detector design.

\subsection{Explainability}
\label{sec:results_explain}
A common and justified concern with CNNs is their tendency to be used as a black box (inputs in, outputs out) with no understanding of their inner working. For detailed physics analyses, this can have significant confidence implications for the final results. Although difficult quantitatively, qualitative assessments of the trained networks can go a long way to prove they behave as desired.

\begin{figure} 
    \centering
    \includegraphics[height=16cm]{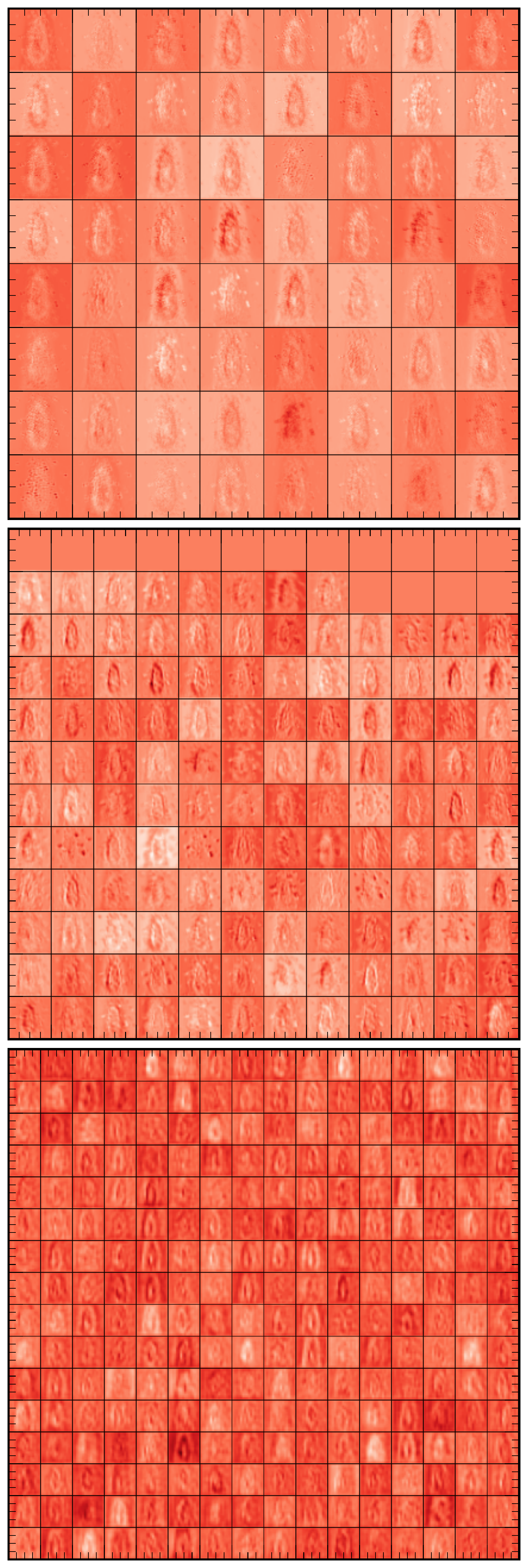}
    \caption{Visualisations of the feature map outputs for the beam classification network form. Shown are all the activated feature map outputs from the first (top), the second (middle), and the third (bottom) VGG blocks with each small box representing an individual feature map. A chipsnet architecture with only a single branch and all three input event maps stacked into a single three-channel input image is used for simplicity.}
    \label{fig:cnn_visualisations}
\end{figure}

Using an example $\nu_{\mu}$ event, visualisations of the output feature maps from the first, second, and third VGG blocks for the trained beam classification network form are shown in Figure~\ref{fig:cnn_visualisations}. Learnt Cherenkov ring features are observed: ring edges, ring holes, outlying hits, Hough peaks, and a myriad of combinations are seen.

Another technique to analyse trained CNNs is t-Distributed Stochastic Neighbour Embedding (t-SNE)~\cite{maaten2008}. The t-SNE procedure is an unsupervised learning algorithm to visualise the learnt high-dimensional feature-space of a trained network in a lower number of dimensions. It accomplishes this by clustering events with similar features nearby in two-dimensional space and separating events with dissimilar features. Here, the outputs from the last fully connected layer before the output layer (with 512 dimensions) are used as input, as they provide the final representation of the learnt network features.

\begin{figure} 
    \includegraphics[width=\textwidth]{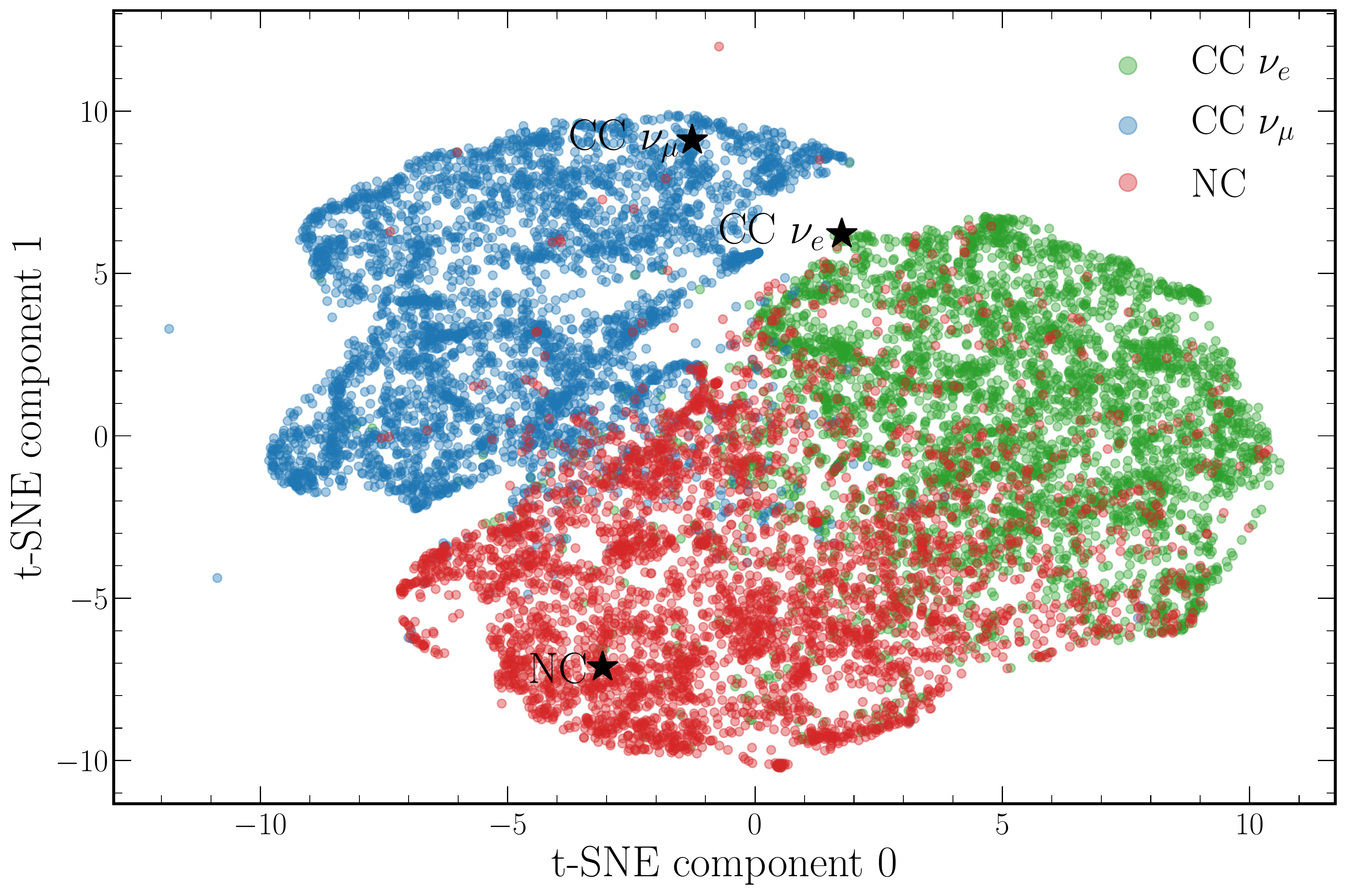}
    \caption{Two dimensional probability space of different beam events generated using the t-SNE procedure on the final fully-connected layer of the trained beam classification network. Three events, one highly CC $\nu_{e}$ like, one highly CC $\nu_{\mu}$ like, and one highly NC like are highlighted and shown in Figure~\ref{fig:explain_beam_tsne_events}.}
    \label{fig:explain_beam_tsne}
\end{figure}

\begin{figure} 
    \includegraphics[width=\textwidth]{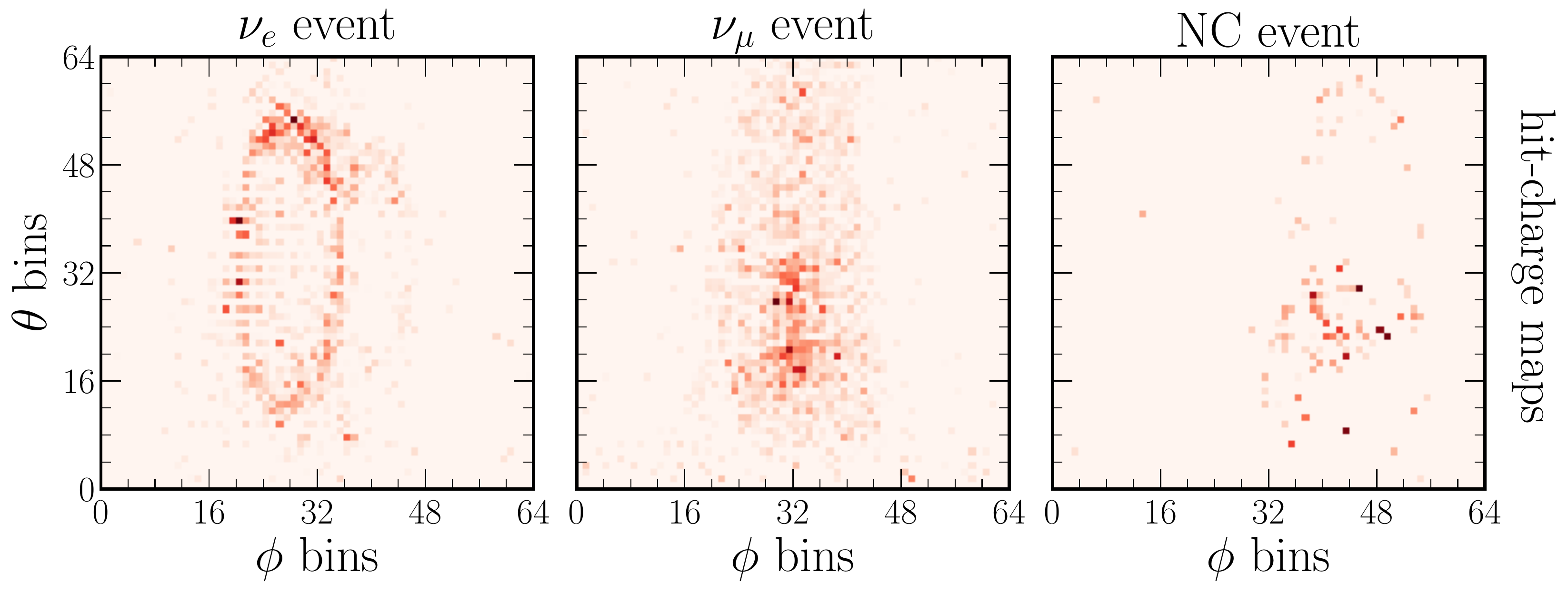}
    \caption{Hit-charge maps of the highly CC $\nu_{e}$ like, CC $\nu_{\mu}$ like, and NC like events from Figure~\ref{fig:explain_beam_tsne}.}
    \label{fig:explain_beam_tsne_events}
\end{figure}

For the beam classification network form, three events, labelled in the t-SNE space of Figure~\ref{fig:explain_beam_tsne}, are shown in Figure~\ref{fig:explain_beam_tsne_events}. Each event is highly representative of its class, achieving a high respective \emph{combined category} score. Both the CC $\nu_{e}$ and NC events are typical of that expected. However, the CC $\nu_{\mu}$ event contains a primary charged lepton that escapes the detector volume, identified by the central peak. This topology suggests that strongly classified CC $\nu_{\mu}$ events can be identified by this `escaping' feature rather than the shape of the muon ring. Future work, therefore, should explore using only fully contained events during beam classification training.

\section{Conclusions and future work}
\label{sec:conclusion}

This paper presented a neutrino event characterisation methodology for the \chips water Cherenkov neutrino detector R\&D project. \chips puts forward a novel water Cherenkov based concept to counter the vast expense, increased complexity, and long construction time expected from future long-baseline neutrino oscillation experiments.

As is the case within the world around us, future \chips concept detectors will also make ever-increasing use of modern machine learning techniques. This comes as a direct result of the dramatic improvement in \chipsfive reconstruction and classification performance brought about by the work presented in this paper. Three forms of a baseline Convolutional Neural Network have been trained to reject cosmic muon events, classify beam events, and estimate neutrino energies, all using only the raw detector event as input. This new approach replaces the standard likelihood-based reconstruction and simple neural network classification, whilst greatly increasing generalisability and processing speed.

The new CNN-based approach is found to provide excellent performance in selecting an efficient and pure appeared CC $\nu_{e}$ sample for which the neutrino energy can be accurately determined. In some cases, the performance is comparable with similar experiments, which is impressive given the significant differences in detector design. The vast cosmic muon background of \chipsfive is found to be accepted by only a factor of $<2.9\pm0.3\times10^{-6}$ (without the help of a veto), equivalent to $<6\pm0.6$ cosmic muon events contaminating the beam sample per year, of which none are expected to be classified as CC $\nu_{e}$ events. 

Furthermore, the key performance metrics for both the CC $\nu_{e}$ and CC $\nu_{\mu}$ beam selections are summarised in Table~\ref{tab:final_metrics}. Not only are the trained CNNs found to provide excellent performance, but some insight into their inner workings was achieved. Cherenkov ring and Hough peak features are extracted from the input images, resulting in a learnt representation of the inputs seen to have strong discriminating power between categories when visualised using the t-SNE technique.

\begin{table}
    \begin{tabular}{lrrr}
        Selection           & Signal Efficiency & Signal Purity & $\sim\nu$ Energy Resolution \\
        \midrule
        CC $\nu_{e}$     & $73.4\pm0.2\%$ & $70.9\pm0.6\%$ & $10.2\pm0.2\%$ \\
        CC $\nu_{\mu}$   & $37.0\pm0.1\%$ & $96.0\pm0.1\%$ & $12.5\pm0.2\%$ \\
    \end{tabular}
    \caption[Key performance metrics of the new CNN approach]
    {The key performance metrics for both the CC $\nu_{e}$ and CC $\nu_{\mu}$ beam selections
        using the new CNN-based approach. The signal efficiency relative to the total number of
        expected events, the signal purity defined as the fraction of selected events which are
        signal, and the approximate signal neutrino energy resolution. The values considering both
        the appeared and beam CC $\nu_{e}$ components as signal are given for the CC $\nu_{e}$
        selection.}
    \label{tab:final_metrics}
\end{table}

It is sincerely hoped that other water Cherenkov neutrino experiments will take inspiration from and then build upon the work presented in this paper for their own Convolutional Neural Network implementations. Although the results presented in this work are somewhat compelling, there are still clear avenues for exploration and improvement. These are all principally related to the critical performance drivers outlined within this paper. 

Firstly, generating the input event maps to focus on the underlying \linebreak Cherenkov profiles is incredibly important; therefore, any methodology to remove distortions further or more accurately determine the interaction vertex position will be beneficial. Secondly, the distribution of events (in energy or type) used within the training sample heavily impacts performance; thus, a comprehensive study of this behaviour could optimise the sample used. Finally, multi-task learning clearly shows promise, with further trial-and-error or a more generalised approach likely to uncover additional valuable tasks. 



\Urlmuskip=0mu plus 1mu\relax
\bibliographystyle{elsarticle-num} 
\bibliography{refs}

\begin{thebibliography}{10}
\expandafter\ifx\csname url\endcsname\relax
  \def\url#1{\texttt{#1}}\fi
\expandafter\ifx\csname urlprefix\endcsname\relax\def\urlprefix{URL }\fi
\expandafter\ifx\csname href\endcsname\relax
  \def\href#1#2{#2} \def\path#1{#1}\fi

\bibitem{adamson2013}
P.~Adamson, et~al., {CHerenkov detectors In mine PitS (CHIPS) letter of intent
  to FNAL}, Preprint (2013).
\newblock \href {https://doi.org/10.48550/arXiv.1307.5918}
  {\path{doi:10.48550/arXiv.1307.5918}}.

\bibitem{psihas2020}
F.~Psihas, M.~Groh, C.~Tunnell, K.~Warburton, A review on machine learning for
  neutrino experiments, International Journal of Modern Physics A 35~(33)
  (2020) 2043005.
\newblock \href {https://doi.org/10.1142/S0217751X20430058}
  {\path{doi:10.1142/S0217751X20430058}}.

\bibitem{fukushima1982}
K.~Fukushima, {Neocognitron: A self-organizing neural network model for a
  mechanism of pattern recognition unaffected by shift in position}, Biological
  Cybernetics (1980) 193--202\href {https://doi.org/10.1007/BF00344251}
  {\path{doi:10.1007/BF00344251}}.

\bibitem{goodfellow2016}
I.~Goodfellow, Y.~Bengio, A.~Courville, {Deep Learning}, MIT Press, 2016,
  \url{http://www.deeplearningbook.org}.

\bibitem{aurisano2016}
A.~Aurisano, A.~Radovic, D.~Rocco, A.~Himmel, M.~Messier, E.~Niner,
  G.~Pawloski, F.~Psihas, A.~Sousa, P.~Vahle, {A convolutional neural network
  neutrino event classifier}, Journal of Instrumentation 11~(09) (2016)
  P09001–P09001.
\newblock \href {https://doi.org/10.1088/1748-0221/11/09/p09001}
  {\path{doi:10.1088/1748-0221/11/09/p09001}}.

\bibitem{szegedy2015}
C.~Szegedy, W.~Liu, Y.~Jia, P.~Sermanet, S.~Reed, D.~Anguelov, D.~Erhan,
  V.~Vanhoucke, A.~Rabinovich, {Going deeper with convolutions}, in:
  {Proceedings of the IEEE conference on computer vision and pattern
  recognition}, 2015, pp. 1--9.
\newblock \href {https://doi.org/10.1109/CVPR.2015.7298594}
  {\path{doi:10.1109/CVPR.2015.7298594}}.

\bibitem{psihas2019}
F.~Psihas, E.~Niner, M.~Groh, R.~Murphy, A.~Aurisano, A.~Himmel, K.~Lang, M.~D.
  Messier, A.~Radovic, A.~Sousa, {Context-enriched identification of particles
  with a convolutional network for neutrino events}, Physical Review D 100~(7)
  (10 2019).
\newblock \href {https://doi.org/10.1103/physrevd.100.073005}
  {\path{doi:10.1103/physrevd.100.073005}}.

\bibitem{baldi2019}
P.~Baldi, J.~Bian, L.~Hertel, L.~Li, {Improved energy reconstruction in NOvA
  with regression convolutional neural networks}, Physical Review D 99~(1) (1
  2019).
\newblock \href {https://doi.org/10.1103/physrevd.99.012011}
  {\path{doi:10.1103/physrevd.99.012011}}.

\bibitem{kronmueller2019}
M.~Kronmueller, T.~Glauch, Application of deep neural networks to event type
  classification in icecube (2019).
\newblock \href {http://arxiv.org/abs/1908.08763} {\path{arXiv:1908.08763}}.

\bibitem{acciarri2017_ref}
R.~Acciarri, et~al., {Design and construction of the MicroBooNE detector},
  Journal of Instrumentation 12~(02) (2017) P02017.
\newblock \href {https://doi.org/10.1088/1748-0221/12/02/P02017}
  {\path{doi:10.1088/1748-0221/12/02/P02017}}.

\bibitem{abratenko2021}
P.~e.~a. Abratenko,
  \href{http://dx.doi.org/10.1103/PhysRevD.103.092003}{Convolutional neural
  network for multiple particle identification in the microboone liquid argon
  time projection chamber}, Physical Review D 103~(9) (May 2021).
\newblock \href {https://doi.org/10.1103/physrevd.103.092003}
  {\path{doi:10.1103/physrevd.103.092003}}.
\newline\urlprefix\url{http://dx.doi.org/10.1103/PhysRevD.103.092003}

\bibitem{acciarri2017}
R.~Acciarri, et~al., {Convolutional neural networks applied to neutrino Events
  in a liquid argon time projection chamber}, Journal of Instrumentation
  12~(03) (2017) P03011–P03011.
\newblock \href {https://doi.org/10.1088/1748-0221/12/03/p03011}
  {\path{doi:10.1088/1748-0221/12/03/p03011}}.

\bibitem{collaboration2020}
B.~Abi, et~al., {Neutrino interaction classification with a convolutional
  neural network in the DUNE far detector} (2020).
\newblock \href {http://arxiv.org/abs/2006.15052} {\path{arXiv:2006.15052}}.

\bibitem{abi2020}
B.~Abi, et~al., {Deep Underground Neutrino Experiment (DUNE), far detector
  technical design report, Volume II DUNE physics} (2020).
\newblock \href {http://arxiv.org/abs/2002.03005} {\path{arXiv:2002.03005}}.

\bibitem{liu2020deep}
J.~Liu, J.~Ott, J.~Collado, B.~Jargowsky, W.~Wu, J.~Bian, P.~Baldi,
  Deep-learning-based kinematic reconstruction for dune, arXiv preprint
  arXiv:2012.06181 (2020).

\bibitem{Rossi2022}
M.~Rossi, S.~Vallecorsa,
  \href{http://dx.doi.org/10.1007/s41781-021-00077-9}{Deep learning strategies
  for protodune raw data denoising}, Computing and Software for Big Science
  6~(1) (Jan 2022).
\newblock \href {https://doi.org/10.1007/s41781-021-00077-9}
  {\path{doi:10.1007/s41781-021-00077-9}}.
\newline\urlprefix\url{http://dx.doi.org/10.1007/s41781-021-00077-9}

\bibitem{domine2020}
L.~Dominé, K.~Terao,
  \href{http://dx.doi.org/10.1103/PhysRevD.102.012005}{Scalable deep
  convolutional neural networks for sparse, locally dense liquid argon time
  projection chamber data}, Physical Review D 102~(1) (Jul 2020).
\newblock \href {https://doi.org/10.1103/physrevd.102.012005}
  {\path{doi:10.1103/physrevd.102.012005}}.
\newline\urlprefix\url{http://dx.doi.org/10.1103/PhysRevD.102.012005}

\bibitem{carloni2022}
K.~Carloni, N.~Kamp, A.~Schneider, J.~Conrad,
  \href{http://dx.doi.org/10.1088/1748-0221/17/02/P02022}{Convolutional neural
  networks for shower energy prediction in liquid argon time projection
  chambers}, Journal of Instrumentation 17~(02) (2022) P02022.
\newblock \href {https://doi.org/10.1088/1748-0221/17/02/p02022}
  {\path{doi:10.1088/1748-0221/17/02/p02022}}.
\newline\urlprefix\url{http://dx.doi.org/10.1088/1748-0221/17/02/P02022}

\bibitem{racah2016}
E.~Racah, S.~Ko, P.~Sadowski, W.~Bhimji, C.~Tull, S.-Y. Oh, P.~Baldi, Prabhat,
  {Revealing fundamental physics from the Daya Bay neutrino experiment using
  deep neural networks}, 2016 15th IEEE International Conference on Machine
  Learning and Applications (ICMLA) (12 2016).
\newblock \href {https://doi.org/10.1109/icmla.2016.0160}
  {\path{doi:10.1109/icmla.2016.0160}}.

\bibitem{aiello2020}
S.~Aiello, et~al., {Event reconstruction for KM3NeT/ORCA using convolutional
  neural networks} (4 2020).
\newblock \href {http://arxiv.org/abs/2004.08254} {\path{arXiv:2004.08254}}.

\bibitem{chipssim2020}
{Chips WCSIM git repository}, \url{https://gitlab.com/chipsneutrino/chips-sim}.

\bibitem{wcsim2020}
{WCSim git repository}, \url{https://github.com/WCSim/WCSim}.

\bibitem{agostinelli2003}
S.~Agostinelli, et~al.,
  \href{http://www.sciencedirect.com/science/article/pii/S0168900203013688}{{Geant4
  -- a simulation toolkit}}, Nuclear Instruments and Methods in Physics
  Research Section A: Accelerators, Spectrometers, Detectors and Associated
  Equipment 506~(3) (2003) 250 -- 303.
\newblock \href {https://doi.org/https://doi.org/10.1016/S0168-9002(03)01368-8}
  {\path{doi:https://doi.org/10.1016/S0168-9002(03)01368-8}}.
\newline\urlprefix\url{http://www.sciencedirect.com/science/article/pii/S0168900203013688}

\bibitem{allison2006}
J.~{Allison}, et~al., {Geant4 developments and applications}, IEEE Transactions
  on Nuclear Science 53~(1) (2006) 270--278.
\newblock \href {https://doi.org/10.1109/TNS.2006.869826}
  {\path{doi:10.1109/TNS.2006.869826}}.

\bibitem{allison2016}
J.~Allison, et~al.,
  \href{http://www.sciencedirect.com/science/article/pii/S0168900216306957}{{Recent
  developments in Geant4}}, Nuclear Instruments and Methods in Physics Research
  Section A: Accelerators, Spectrometers, Detectors and Associated Equipment
  835 (2016) 186 -- 225.
\newblock \href {https://doi.org/https://doi.org/10.1016/j.nima.2016.06.125}
  {\path{doi:https://doi.org/10.1016/j.nima.2016.06.125}}.
\newline\urlprefix\url{http://www.sciencedirect.com/science/article/pii/S0168900216306957}

\bibitem{acciarri2016}
R.~Acciarri, et~al., {Long-Baseline Neutrino Facility (LBNF) and Deep
  Underground Neutrino Experiment (DUNE) conceptual design report volume 1: the
  LBNF and DUNE projects} (2016).
\newblock \href {http://arxiv.org/abs/1601.05471} {\path{arXiv:1601.05471}}.

\bibitem{campbell2020}
M.~Campbell, {Measuring neutrino oscillations in the NOvA and CHIPS Detectors},
  Ph.D. thesis, University College London,
  \url{https://discovery.ucl.ac.uk/id/eprint/10097512/1/Campbell_10097512_thesis_sig-removed.pdf}
  (8 2020).

\bibitem{hamamatsu_handbook}
{Photomultiplier Tubes - Basics and Applications},
  \url{https://www.hamamatsu.com/resources/pdf/etd/PMT_handbook_v3aE.pdf}.

\bibitem{andreopoulos2009}
C.~Andreopoulos, et~al., {The GENIE neutrino Monte Carlo generator}, Nucl.
  Instrum. Meth. A 614 (2010) 87--104.
\newblock \href {http://arxiv.org/abs/0905.2517} {\path{arXiv:0905.2517}},
  \href {https://doi.org/10.1016/j.nima.2009.12.009}
  {\path{doi:10.1016/j.nima.2009.12.009}}.

\bibitem{andreopoulos2015}
C.~Andreopoulos, C.~Barry, S.~Dytman, H.~Gallagher, T.~Golan, R.~Hatcher,
  G.~Perdue, J.~Yarba, {The GENIE neutrino Monte Carlo generator: physics and
  user manual} (10 2015).
\newblock \href {http://arxiv.org/abs/1510.05494} {\path{arXiv:1510.05494}}.

\bibitem{hagmann2012_1}
C.~Hagmann, D.~Lange, J.~Verbeke, D.~Wright,
  \href{https://nuclear.llnl.gov/simulation/doc_cry_v1.7/cry.pdf}{{Cosmic-ray
  shower library (CRY)}} (3 2012).
\newline\urlprefix\url{https://nuclear.llnl.gov/simulation/doc_cry_v1.7/cry.pdf}

\bibitem{hagmann2012_2}
C.~Hagmann, D.~Lange, D.~Wright,
  \href{https://nuclear.llnl.gov/simulation/doc_cry_v1.7/cry_physics.pdf}{{Monte
  Carlo simulation of proton-induced cosmic-ray cascades in the atmosphere}} (2
  2012).
\newline\urlprefix\url{https://nuclear.llnl.gov/simulation/doc_cry_v1.7/cry_physics.pdf}

\bibitem{klimushin2001}
S.~Klimushin, E.~Bugaev, I.~A. Sokalski, {Precise parametrizations of muon
  energy losses in water}, in: 27th International Cosmic Ray Conference,
  Vol.~3, 2001, p. 1009.
\newblock \href {http://arxiv.org/abs/hep-ph/0106010}
  {\path{arXiv:hep-ph/0106010}}.

\bibitem{tutorial2020}
{WCSim: How to add your own photodetector tutorial},
  \url{https://github.com/WCSim/WCSim/wiki/Tutorial:-How-to-add-your-own-photdetector}.

\bibitem{chipsreco2020}
{Chips reconstruction git repository},
  \url{https://gitlab.com/chipsneutrino/chips-reco}.

\bibitem{alzubaidi2021review}
L.~Alzubaidi, J.~Zhang, A.~J. Humaidi, A.~Al-Dujaili, Y.~Duan, O.~Al-Shamma,
  J.~Santamar{\'\i}a, M.~A. Fadhel, M.~Al-Amidie, L.~Farhan, Review of deep
  learning: Concepts, cnn architectures, challenges, applications, future
  directions, Journal of big Data 8~(1) (2021) 1--74.

\bibitem{chipsnet2020}
{chipsnet git repository}, \url{https://gitlab.com/chipsneutrino/chips-net}.

\bibitem{tf2015}
M.~Abadi, et~al., \href{https://www.tensorflow.org/}{{TensorFlow: large-scale
  machine learning on heterogeneous systems}}, software available from
  \url{https://www.tensorflow.org/} (2015).
\newline\urlprefix\url{https://www.tensorflow.org/}

\bibitem{theodore2016}
T.~Theodore, {Particle identification in Cherenkov detectors using
  convolutional neural networks}, Master's thesis, University of Toronto
  (2016).

\bibitem{illingworth1988}
J.~Illingworth, J.~Kittler, A survey of the hough transform, Computer vision,
  graphics, and image processing 44~(1) (1988) 87--116.

\bibitem{simonyan2014}
K.~Simonyan, A.~Zisserman, {Very deep convolutional networks for large-scale
  image recognition} (2014).
\newblock \href {http://arxiv.org/abs/1409.1556} {\path{arXiv:1409.1556}}.

\bibitem{ioffe2015}
S.~Ioffe, C.~Szegedy, {Batch Normalization: accelerating deep network training
  by reducing internal covariate shift}, in: {Proceedings of the 32nd
  International Conference on International Conference on Machine Learning -
  Volume 37}, ICML'15, JMLR.org, 2015, p. 448–456.
\newblock \href {http://arxiv.org/abs/1502.03167} {\path{arXiv:1502.03167}}.

\bibitem{hu2018}
J.~Hu, L.~Shen, G.~Sun, {Squeeze-and-excitation networks}, in: {Proceedings of
  the IEEE conference on computer vision and pattern recognition}, 2018, pp.
  7132--7141.
\newblock \href {https://doi.org/10.1109/TPAMI.2019.2913372}
  {\path{doi:10.1109/TPAMI.2019.2913372}}.

\bibitem{chollet2015}
F.~Chollet, et~al., {Keras}, \url{https://keras.io} (2015).

\bibitem{kendall2018}
A.~Kendall, Y.~Gal, R.~Cipolla, {Multi-task learning using uncertainty to weigh
  losses for scene geometry and semantics}, in: {Proceedings of the IEEE
  conference on computer vision and pattern recognition}, 2018, pp. 7482--7491.
\newblock \href {https://doi.org/10.1109/CVPR.2018.00781}
  {\path{doi:10.1109/CVPR.2018.00781}}.

\bibitem{kingma2014}
D.~P. Kingma, J.~Ba, {Adam: a method for stochastic optimization} (2014).
\newblock \href {http://arxiv.org/abs/1412.6980} {\path{arXiv:1412.6980}}.

\bibitem{hertel2020}
L.~Hertel, J.~Collado, P.~Sadowski, J.~Ott, P.~Baldi, {Sherpa: robust
  hyperparameter optimization for machine learning} (2020).
\newblock \href {http://arxiv.org/abs/2005.04048} {\path{arXiv:2005.04048}}.

\bibitem{esteban2020}
I.~Esteban, M.~Gonzalez-Garcia, M.~Maltoni, T.~Schwetz, A.~Zhou, {The fate of
  hints: updated global analysis of three-flavor neutrino oscillations} (7
  2020).
\newblock \href {http://arxiv.org/abs/2007.14792} {\path{arXiv:2007.14792}}.

\bibitem{list2002}
B.~List, {Why and when to optimize efficiency times purity}, ETH Zurich
  internal note (2002).

\bibitem{jtingey2021}
J.~Tingey, {Convolutional neural networks for the CHIPS neutrino detector R\&D
  project}, Ph.D. thesis, University College London,
  \url{https://discovery.ucl.ac.uk/id/eprint/10129514/1/thesis.pdf} (5 2021).

\bibitem{blake2016}
A.~Blake, S.~Germani, Y.~B. Pan, A.~J. Perch, M.~M. Pfützner, J.~Thomas, L.~H.
  Whitehead, {CHIPS event reconstruction and design optimisation} (2016).
\newblock \href {http://arxiv.org/abs/1612.04604} {\path{arXiv:1612.04604}}.

\bibitem{acero2019}
M.~A. Acero, et~al., {First measurement of neutrino oscillation parameters
  using neutrinos and antineutrinos by NOvA}, Physical Review Letters 123
  (2019) 151803.
\newblock \href {https://doi.org/10.1103/PhysRevLett.123.151803}
  {\path{doi:10.1103/PhysRevLett.123.151803}}.

\bibitem{abe2015}
K.~Abe, et~al., {Measurements of neutrino oscillation in appearance and
  disappearance channels by the T2K experiment with $6.6\times 10^{20}$ protons
  on target}, Phys. Rev. D 91 (2015) 072010.
\newblock \href {https://doi.org/10.1103/PhysRevD.91.072010}
  {\path{doi:10.1103/PhysRevD.91.072010}}.

\bibitem{jiang2019}
M.~Jiang, et~al., {Atmospheric neutrino oscillation analysis with improved
  event reconstruction in Super-Kamiokande IV}, Progress of Theoretical and
  Experimental Physics 2019~(5), 053F01 (05 2019).
\newblock \href
  {http://arxiv.org/abs/https://academic.oup.com/ptep/article-pdf/2019/5/053F01/28638877/ptz015.pdf}
  {\path{arXiv:https://academic.oup.com/ptep/article-pdf/2019/5/053F01/28638877/ptz015.pdf}},
  \href {https://doi.org/10.1093/ptep/ptz015} {\path{doi:10.1093/ptep/ptz015}}.

\bibitem{maaten2008}
L.~v.~d. Maaten, G.~Hinton,
  \href{http://www.jmlr.org/papers/volume9/vandermaaten08a/vandermaaten08a.pdf}{{Visualizing
  data using t-SNE}}, {Journal of Machine Learning Research} 9~(Nov) (2008)
  2579--2605.
\newline\urlprefix\url{http://www.jmlr.org/papers/volume9/vandermaaten08a/vandermaaten08a.pdf}

\end{thebibliography}





\end{document}